\documentclass[reprint,aps,prb,twocolumn,amsmath,amssymb]{revtex4-2}

\usepackage{amsmath}
\usepackage{amssymb}
\usepackage{graphicx}
\usepackage{verbatim}
\usepackage{subfigure}
\usepackage{hyperref}
\usepackage{natbib}
\usepackage{physics}
\usepackage{caption}
\usepackage{bbold}
\usepackage{float}
\usepackage[english]{babel}
\usepackage{xcolor}
\usepackage[normalem]{ulem}
\usepackage{cancel}

\newcommand\id{\mathbb{1}}
\newcommand\up{\uparrow}
\newcommand\down{\downarrow}

\begin{document}

\title{Localization Dynamics from static and mobile Impurities}

\author{Ephraim Bernhardt, Fan Yang and Karyn Le Hur}
\affiliation{CPHT, CNRS, \'Ecole Polytechnique, Institut Polytechnique de Paris, Route de Saclay, 91128 Palaiseau, France}

\begin{abstract}
We study the superfluid response and localization dynamics from static and mobile impurities. The superfluidity is formed in the rung-Mott phase of a bosonic ladder model producing spin-Meissner
currents induced by a $\mathbb{U}(1)$ gauge field or a uniform magnetic field. Impurities are described through two-state systems which act as a two-peak random potential. An impurity sits either at the top or at the bottom of the ladder on each rung equally, producing a telegraph signal. The impurities-matter coupling gives rise to a classical Ising symmetry for static and mobile impurities associated to the inversion symmetry of the two legs of the ladder. From the decoupled rungs limit, we also identify a local $\mathbb{Z}_2$ gauge theory for mobile impurities. The properties of the system are studied from an effective quantum spin model including the possibility of four-body coupling in the limit of a strong interaction between bosons and impurities. Through analytical approaches and numerical exact diagonalization, we study the superfluid currents both in the weakly-coupled and strongly-coupled rungs limits for the bosons. In the weakly-coupled rungs situation, we find a smooth power-law localization whereas the strongly-coupled rungs limit produces a steep localization or insulating phase for various configurations of the two-peak random potential. In the strongly disordered situation, through entanglement and bipartite fluctuation measures, we also identify a many-body localization regime in time after a quench of the system when prepared in a N\' eel state.
\end{abstract}

\maketitle
\date{\today}

\setcounter{secnumdepth}{3}
\thispagestyle{empty}
\newpage
\pagenumbering{arabic}

\newpage

\section{Introduction}

Understanding the physical mechanisms related to the occurrence of phases of matter and their transitions is important for fundamental purposes and also for practical applications. In spin systems, it is quite natural to identify $\mathbb{Z}_2$ symmetries. The classical Ising chain
\begin{equation}
 H = J \sum_j \tau_j \tau_{j+1},
\end{equation}
with $\tau_j = \pm 1$ is obviously invariant under a simultaneous transformation $\tau_j \rightarrow -\tau_j$ on all sites. For quantum systems, $\mathbb{Z}_2$ lattice gauge theories have been then developed from the spin algebra and local symmetries in a variety of strongly-correlated systems such as Ising and Kitaev spin models, high-Tc superconductors and light-matter Hamiltonians \cite{wen2004quantum, sachdev2011quantum}. It is important to mention progress on realizations of lattice gauge theories in quantum technology \cite{Pan,Esslinger,MariCarmen,Heidelberg}. In this article, we will start from the $\mathbb{Z}_2$ classical symmetry $\tau_j\rightarrow -\tau_j$. The application of a uniform magnetic field producing a $\mathbb{U(1)}$ gauge field is also known to induce a variety of interesting phases in these one-dimensional ladder systems such as a Meissner-like phase, vortex phase or fractional quantum Hall phases \cite{petrescu2013bosonic, atala2014observation, orignac2001meissner, lehur2018driven}. Therefore, we propose a model combining features of both effects: On the one hand, we will map a bosonic ladder model to a quantum spin chain with $\mathbb{Z}_2$ symmetry. On the other hand, we will include a $\mathbb{U(1)}$ gauge field which allows us to define and study a superfluid current flow in the system. We will then address the role of impurities on the current profile associated to the superflow.

Studying the effect of impurities opens the door to the vast field of research about localization phenomena and disorder effects. The basic question in this area goes back to Anderson, who asked how mobile particles can get localized in the presence of disorder \cite{anderson1958absence}. It turned out that this localization can even occur in the presence of interactions \cite{fisher1989boson,GiamarchiSchulz,gabriel,Toulouse}, leading recently to many-body localization which is an active area of both experimental and theoretical research \cite{alet2018many, abanin2019colloquium}. Recently, the case of mobile impurities in a quantum fluid has been shown to result in interesting resonance phenomena \cite{ThomasPeter} related to the Kane-Fisher double barrier model \cite{KaneFisher}. Related to phases of matter and transitions between them,  disorder and impurities produce interesting physics which is worth being studied through different probes \cite{Munich,googleMBL,cQEDPan}. In turn, the transition between different types of localization is not fully understood, in particular with respect to many-body localization \cite{singh2016signatures}. One of the most striking characteristics of many-body localized phases is that they do not fulfill the eigenstate thermalization hypothesis, meaning that the initial state will manifest itself in all later states of the system \cite{rigol2008thermalization, huse2013localization,abanin2019colloquium}. This property and transitions have been thoroughly analyzed using bipartite fluctuations and entanglement measures \cite{singh2016signatures,song2012bipartite}. It is important to emphasize here that two-fluids systems in relation with localization effects and gauge theories \cite{Fishing,Brenes} have attracted  attention these last years, in particular through quantum spin models \cite{FreeLoc,FreeLoc2} and ladder models  \cite{Moore}.

In this article, we present conclusions about the localization by directly studying the current along a ladder-shaped lattice. We introduce a second particle species to this lattice and distinguish the cases where these behave as a static disorder thus forming a telegraph signal and where impurities correspond to dynamical two-state or spin-1/2 systems. We address the correlated limit for the bosonic particles referring to the rung-Mott phase with one particle (boson) per rung such that the system is in fact a spin superfluid when applying a uniform magnetic field \cite{petrescu2013bosonic} and such that the total Hamiltonian can be re-written as a quantum spin model. The occurrence of a spin Meissner current, which represents here the physical response to the applied magnetic field or $\mathbb{U}(1)$ gauge field, is stabilized from the Josephson effect between wires. The bosons-impurities coupling has an intrinsic $\mathbb{Z}_2$ symmetry similar to the classical Ising model. This is equivalent to the sub-lattice symmetry in our case corresponding to invert the two legs of the ladder. In the limit where all the rungs are decoupled, we also identify a relation with local $\mathbb{Z}_2$ quantum gauge theories in the presence of a magnetic flux. This will allow us to study the interplay between the $\mathbb{U(1)}$ and $\mathbb{Z}_2$ gauge fields  which become dynamical. Including a finite coupling between the rungs, the global $\mathbb{Z}_2$ gauge symmetry present in Refs. \cite{schweizer2019floquet, barbiero2018coupling} is explicitly broken by the magnetic field but the sub-lattice symmetry remains. As a direct comparison, we study the evolution of observables from the weakly coupled rungs limit. We will also describe the effect of the telegraph potential on the persistent current regime for small Josephson coupling between the legs of the ladder. Finally, we will show many-body localization physics in time when preparing the rung-Mott phase in an antiferromagnetic N\' eel state along the $z$-axis.

The ladder geometry is frequently being used in cold atom experiments and allows to implement both $\mathbb{U}(1)$ and $\mathbb{Z}_2$ gauge fields \cite{atala2014observation, schweizer2019floquet, barbiero2018coupling}. Through an appropriate periodic driving protocol, both types of symmetries can be realized \cite{barbiero2018coupling, goldman2014periodically, goldman2015periodically}.  It has been suggested that it can also be useful to test localization effects \cite{crepin2011phase, ye2019propagation, carrasquilla2011bose}. The progress in experimental techniques using Bose-Einstein condensates in optical lattices also allows to address Mott physics \cite{jaksch1998cold, greiner2002quantum, song2021realizing}. We emphasize here that the spin superfluid current in the rung-Mott phase has not been studied in Ref. \cite{barbiero2018coupling} in relation with $\mathbb{Z}_2$ gauge theories. We also note other recent theoretical studies with a telegraph potential or binary disorder showing feasible applications of many-body localization in the ladder systems \cite{andraschko2014purification, tang2015quantum} which considered a ladder without Josephson coupling between the wires.

\par The article is organized as follows. In Sec. \ref{TheModel}, we introduce the model and remind definitions related to the rung-Mott state of the bosonic ladder system \cite{petrescu2013bosonic}. In Sec. \ref{runglimit}, we show distinct limits of solvable quantum dynamics in the presence of impurities associated to a smooth localization of the (spin) Meissner superfluid response, with a power-law profile, in the situation of {\it weakly-coupled rungs}. We will first study the case of static impurities behaving then as purely classical objects commuting with the Hamiltonian and then the case where impurities acquire a quantum dynamics. In the latter case, we will show a connection with a one-dimensional $\mathbb{Z}_2$ lattice gauge theory from the limit of decoupled rungs. Four-body spin Hamiltonians occur in the limit of strong interactions with impurities. In Sec. \ref{stronginteractions}, we describe the limit of {\it strong interrung} interactions when including a prominent longitudinal hopping term for the bosons and study the behavior of the persistent current. We make a bridge with fermions through the Jordan-Wigner transformation in the case of the telegraph signal when averaging over different configurations. This shows the occurrence of an insulating (localized) phase from the profile of the current at strong coupling with the impurities. We also address the specific case of aligned impurities. For the situation of antiferromagnetically aligned impurities, we apply the bosonization formalism of Luttinger liquids \cite{haldane1981luttinger} and renormalization group arguments. We observe a strong (steep) localization effect and compare the results with Gaussian disorder \cite{giamarchi2003quantum}. In Sec. \ref{MBL}, we discuss the limit of many-body localization from quantum spin chain models \cite{luitz2015many, singh2016signatures}, including here the particular profile of the telegraph potential. To realize this limit, the system evolves as a mixed state produced after a quench when preparing the rung-Mott system in a N\' eel state. We study the time-dependent profile of entanglement and bipartite fluctuation measures \cite{singh2016signatures,song2012bipartite}.

We present various analytical methods and compare with results from numerical exact diagonalizations (ED). Information on the numerical approach is shown in Appendix \ref{numerics}. In Appendix \ref{bloch_sphere_calculation}, we discuss mathematical derivations related to the Bloch sphere theory in Sec. \ref{static_impurities}. Details on the calculation of observables with mobile impurities discussed in Sec. \ref{mobile} are given in Appendix \ref{mobile_impurities_calculation}. In Appendix \ref{fourbody_derivation} we derive a four-body spin Hamiltonian related to Sec. \ref{large_impurity_interaction}.

\section{The model}
\label{TheModel}

\begin{figure}[b]\centering
  \includegraphics[width=0.5\textwidth]{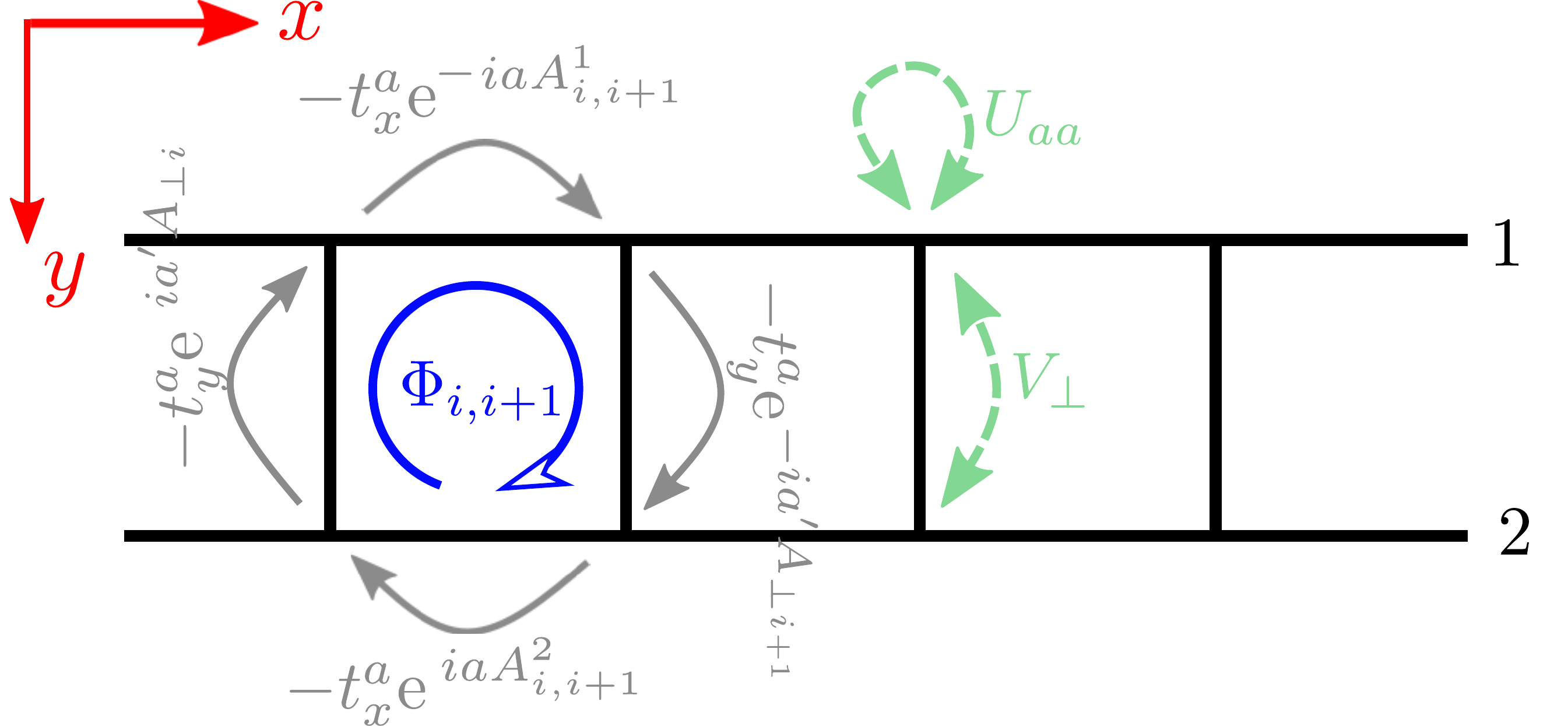}
  \caption{(color online) The setup of a bosonic ladder with hopping strengths (in gray) and potentials felt by the $a$-particles (in green). The effect of a magnetic vector potential enters through the Peierls substitution as complex phases to the hopping strengths. Going around one square plaquette, a flux
  $\Phi_{i,i+1}$ is acquired.}
	\label{first_setup}
\end{figure}

We introduce the bosonic two-leg ladder populated by a particle species which we will in the following often call {\it `$a$-particles'} \cite{petrescu2013bosonic, petrescu2015chiral}. In this model, we want to access a variety of different scenarios, for which we introduce the following parameters: First of all, there should be hopping along the legs of the ladder, which we consider as the $x$-direction, with an amplitude $t_x^a$. There is also hopping along the rungs of the ladder, in the $y$-direction, with an amplitude $t_y^a$. Different phases of this model can be controlled by three energy-scales. A chemical potential $\mu$ determines the filling of the ladder. Furthermore, we introduce an on-site repulsion potential $U_{aa}$ penalizing two bosonic particles sitting on the same site. Lastly, we also introduce a potential $V_\perp$ causing repulsion for two particles on the same rung. This setup is shown pictorially in Fig. \ref{first_setup}.

The effect of a magnetic field can be included through a Peierls substitution following \cite{petrescu2013bosonic}.
In a link to cold atom experiments, this can be realized using a large on-site potential and refacilitating the hopping by driving the system periodically in time \cite{goldman2014periodically, goldman2015periodically}.
Then, we add a second particle species also living on this same ladder and following its own dynamics, while interacting with the $a$-particles, inspired from \cite{barbiero2018coupling}.
We will call this second particle species {\it `gauge particles'}, {\it`$f$-particles'} or {\it impurities}, due to the different roles they play in the article.
A priori, we could imagine $a$- and $f$-particles with similar dynamics and influencing each other through an interaction potential. In the following, we will however be mainly interested in the dynamics of the $a$-particles in the presence of the impurities. We therefore assume to have exactly one $f$-particle on each rung, which corresponds to half-filling and an infinitely high on-site and on-rung repulsion. The dynamics of the $f$-particles on the lattice can then be identified  with the telegraph signal. When these impurities are entirely static, this is similar to imposing a kind of quenched disorder on the $a$-particles.

\par
The interaction between the two particle species is realized in a density dependent way with an energy scale $U_{af}$. For clarity, first we address the situation of static impurities in Eq. (\ref{petrescu_hamilton}) below.
In summary, we get the following Hamiltonian for the $a$-particles interacting with the impurities:
\begin{multline}
\label{petrescu_hamilton}
  H = -t^a_x \sum_{\alpha,i} \mathrm{e}^{iaA_{i,i+1}^\alpha} a^{\dag}_{\alpha i} a_{\alpha,i+1 } +\text{h.c.} \\
  - t^a_y \sum_{i} \mathrm{e}^{-ia'A_{\perp i}} a^{\dag}_{2 i} a_{1 i}  +\text{h.c.} + \frac{U_{aa}}{2}\sum_{\alpha,i} n^a_{\alpha i} (n^a_{\alpha i}-1) \\
  + V_{\perp}\sum_{ i} n^a_{1 i} n^a_{2 i}- \mu \sum_{\alpha, i} n^a_{\alpha i} + U_{af} \sum_{\alpha,i} n_{\alpha,i}^a n_{\alpha,i}^f.
\end{multline}

We use the following notation: Superscripts $a$ or $f$ designate the respective particle species. In the sums, $\alpha$ denotes the leg of the ladder ($1$ or $2$, as shown in Fig. \ref{first_setup}), $i$ goes along the rungs. Consequently, $a_{\alpha i}^\dag$ and $a_{\alpha i}$ are the bosonic creation and annihilation operators of an $a$-particle at a rung $i$ and the leg $\alpha = 1,2$ of the ladder, with the number operator $n^a_{\alpha i} = a_{\alpha i}^\dag a_{\alpha i}$.
The number operator related to the $f$-particles $n^f_{\alpha i}$ is defined similarly. The phases of the hopping terms enter through a Peierls substitution for a uniform external magnetic field (or gauge field).
They are different along the legs ($\mathrm{e}^{iaA_{i,i+1}^\alpha}$) and along the rungs ($\mathrm{e}^{ia'A_{\perp i}}$) with $A_{i,i+1}^\alpha$ and $A_{\perp i}$ being the components of the vector potential at the respective link of the ladder and $a$ and $a'$ the respective lattice spacings (see Fig. \ref{first_setup}) \cite{petrescu2013bosonic}. The flux per plaquette can be evaluated by a contour integral around a plaquette and through Stokes theorem shows the following relation with a uniform magnetic field \cite{orignac2001meissner, petrescu2013bosonic}:
\begin{equation}
\label{Phi}
	\Phi_{i,i+1} = \oint \mathbf{A} \cdot d \mathbf{l} = -a (A_{i,i+1}^1 - A_{i,i+1}^2) - a'(A_{\perp i+1} - A_{\perp i}).
\end{equation}

\subsection{Rung-Mott Phase and Definitions}\label{rungMottDefinitions}

To realize the spin model with the $\mathbb{Z}_2$ symmetry, we assume the system to be in the Mott phase for the $a$-particles. The particles are localized with one particle per rung referring to the rung-Mott state. For clarity's sake, here we fix the definitions starting from Ref. \onlinecite{petrescu2013bosonic}. A simple matrix analysis of the system neglecting the interaction with impurities shows that the system is in the Mott phase for $V_\perp + t_y^a > \mu > -t_y^a$. Treating the hopping along the legs of the ladder perturbatively allows to derive the following effective spin Hamiltonian by considering all possible second-order processes:
\begin{widetext}
\begin{eqnarray}\label{effective_spin}
  H = \sum_i -2J_{xy}\mathrm{e}^{iaA^\parallel_{i,i+1}} \sigma_i^+ \sigma_{i+1}^- +\text{h.c.} + J_z \sigma_i^z \sigma_{i+1}^z - g\left(\cos(a'A_{\perp i})\sigma_i^x - \sin(a'A_{\perp i})\sigma_i^y \right) + \frac{U_{af}}{2} \sigma_i^z \tau^z_i.
\end{eqnarray}
\end{widetext}
Here, we define $J_{xy} = (t_x^a)^2/V_\perp$, $J_z = (t_x^a)^2 (-2/U_{aa} + 1/V_\perp)$, $g = t_y^a$ and wrote $A^\parallel_{i,i+1} = A_{i,i+1}^1 - A_{i,i+1}^2$.

The second-order induced terms for the $a$-particles can be identified as a correlated hopping term ($J_{xy}$) between the two wires and an Ising interaction term ($J_z$). Both are tunable when varying the potentials $V_{\perp}$ and $U_{aa}$ which are in principle both positive to obtain a larger region with the rung-Mott phase. We have $J_z=0$ when $V_{\perp}=U_{aa}/2$ and otherwise $J_z$ can take both signs. The particle creation and annihilation operators have been replaced by Pauli matrices $\sigma_i^\alpha$ with $\alpha = x,y,z$ for effective spins due to $a$-particles. More precisely, we used the Schwinger-boson representation of the SU(2) algebra \cite{schwinger1952lectures} with $\sigma^x_i = a_{1i}^\dag a_{2i} + a_{1i} a_{2i}^\dag$, $\sigma^y_i = -i a_{1i}^\dag a_{2i} + i a_{1i} a_{2i}^\dag$ and $\sigma^z_i = a_{1i}^\dag a_{1i} - a_{2i}^\dag a_{2i}$ (and therefore $\sigma_i^+ = a_{1i}^\dag a_{2i}$ and $\sigma_i^- = a_{1i} a_{2i}^\dag$.
In analogy to the Schwinger-boson representation, we defined $\tau_i^z = n_{1i}^f - n_{2i}^f$ for the $f$-particles, which here are to be thought of as classical two-state systems defined through $n_{1i}^f=f_{1i}^\dag f_{1i}$ and $n_{2i}^f=f_{2i}^\dag f_{2i}$.

\begin{figure}[h]
  \centering
  \includegraphics[width=\linewidth]{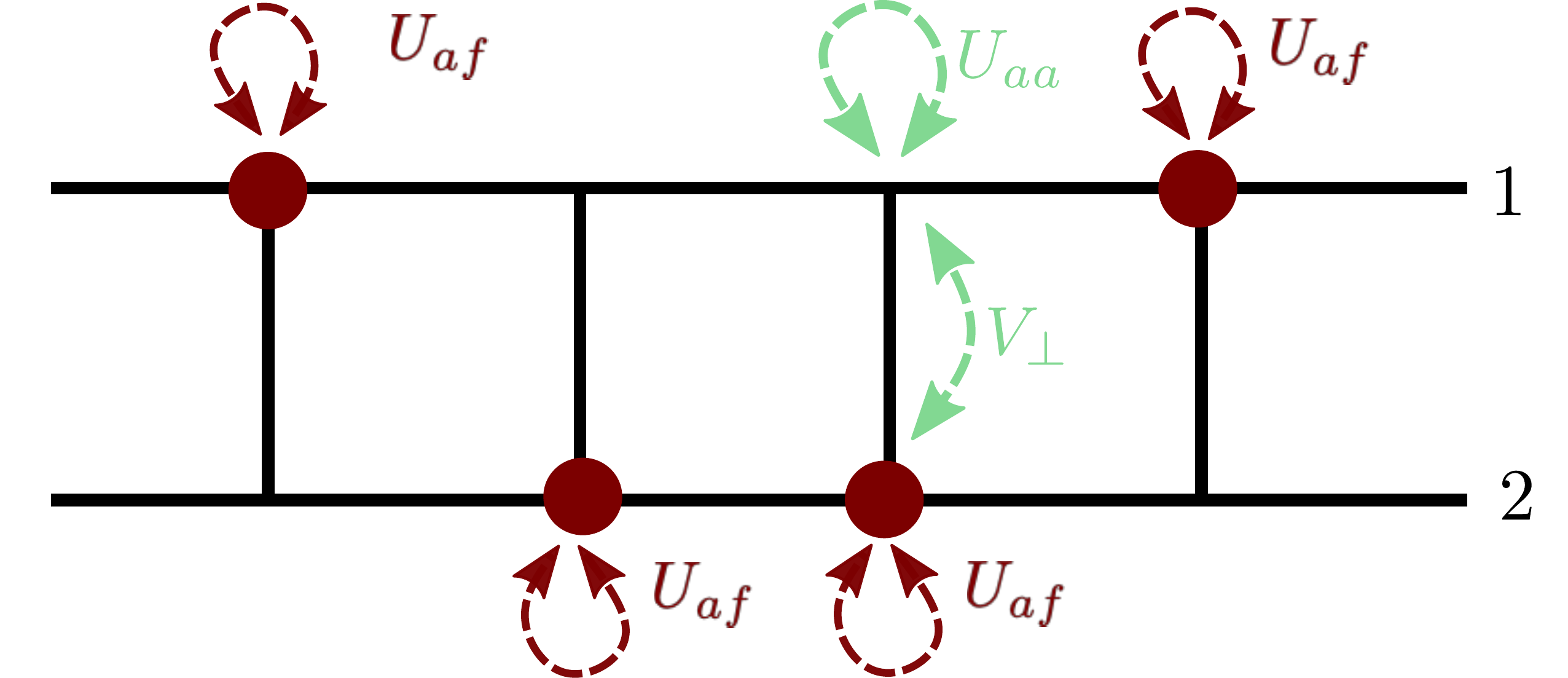}
  \captionof{figure}{The $f$-particles shown in red cause an additional on-site potential felt by the $a$-particles which we call $U_{af}$. This potential is random in a sense that the $f$-particles are at each rung randomly placed on one of the two legs referring then to the telegraph signal.}
	\label{f_configuration}
\end{figure}

Formulating the Hamiltonian using spin operators is justified since there is precisely one particle of each species  on a given rung, as the $a$-particles are in the Mott phase and we assumed that the $f$-particles are two-state systems in a general sense e.g. spinless fermions such that $f^{\dagger}_{1i} f_{1i}+f^{\dagger}_{2i}f_{2i}=1$. This allows us to write their states at each rung in the basis $\ket{10}_i$, $\ket{01}_i$ which can be identified with the spin basis $\ket{\up}_i$, $\ket{\down}_i$.
Without the Peierls phases, the Hamiltonian \eqref{effective_spin} is that of an XXZ-model in a magnetic field, where the transverse field acting on each rung $g$ represents the Josephson term for the $a$-particles.
Here, we also introduce the matter-impurities coupling $U_{af}$ which now represents an Ising coupling between spins. The term $U_{af}$ can be realized through an interaction between two species similar to a Hubbard interaction as shown in Fig. \ref{f_configuration}. The choice of parameters within the rung-Mott phase for the $a$-particles is motivated by the occurrence of a $\mathbb{Z}_2$ (classical) symmetry under flipping all $z$-spin components, i.e. changing $\sigma_i^z \rightarrow -\sigma_i^z$ and $\tau_i^z \rightarrow -\tau_i^z$ simultaneously. Physically, this symmetry corresponds to invert the two legs $1$ and $2$, referring then to the sub-lattice symmetry on a rung $i$, such that $a_{1i}\leftrightarrow a_{2i}$ and $f_{1i}\leftrightarrow f_{2i}$. This also leads to $\sigma_i^y\rightarrow -\sigma_i^y$ such that the $SU(2)$ quantum spin algebra for the $\vec{\sigma}$ spins is preserved.

The $1\leftrightarrow 2$ sub-lattice symmetry can be implemented through the operator
\begin{equation}
\label{R}
{\cal R} = \prod_i \sigma_i^x{\cal O} \otimes \mathbb{Z}_2,
\end{equation}
where the $\mathbb{Z}_2$ symbol defines the classical Ising symmetry $\tau^z_i\rightarrow -\tau^z_i$. In the presence of a magnetic flux, the transformation $1\leftrightarrow 2$ is implemented through the modification $A_{\perp i}\rightarrow -A_{\perp i}$ and $A_{i,i+1}^{\parallel}\rightarrow -A_{i,i+1}^{\parallel}$ such that ${\cal O}f(A_{\perp i},A_{i,i+1}^{\parallel})|\Psi\rangle = f(-A_{\perp i},-A_{i,i+1}^{\parallel}){\cal O}|\Psi\rangle$ for any state $|\Psi\rangle$. The ${\cal O}$ operator corresponds to invert the direction of the magnetic field on each square of the ladder simultaneously such that the ${\cal R}$ symmetry, which commutes with the Hamiltonian, is defined in a gauge-independent form. This sub-lattice symmetry will be present in all the Sections of the article.

In Sec. \ref{mobile} for the situation of mobile impurities, we will also discuss $\mathbb{Z}_2$ quantum gauge theories starting from the observation that in the decoupled rungs limit, with $J_{xy}=J_z=0$, we can also define a local $\mathbb{Z}_2$ symmetry
\begin{equation}
\label{G}
 \mathcal{G}_i = \left(\cos(a'A_{\perp i})\sigma_i^x - \sin(a'A_{\perp i})\sigma_i^y \right)\otimes \mathbb{Z}_2,
\end{equation}
such that $\mathcal{G}_i=H \mathcal{G}_i H^{-1}$. In Sec. \ref{mobile}, the $\mathbb{Z}_2$ symmetry $\tau_i^z\rightarrow -\tau_i^z$ in Eq. \ref{G} will be implemented quantum mechanically through the $\tau_i^x$ operator acting on each rung.
Here, $\mathcal{G}_i$ has a well-defined local origin starting from a double-well limit or one rung that we will study in Sec. \ref{mobile} related to $\mathbb{Z}_2$ gauge theories. We will analyse the consequences of the local $\mathbb{Z}_2$ gauge theory in Eq. (\ref{G}) on physical observables starting from decoupled rungs and mobile impurities bouncing back and forth between the top and bottom sites.
It should be mentioned that in the presence of a finite flux, $\mathcal{G}=\prod_i \mathcal{G}_i$ does not commute with the Hamiltonian when including finite values of $J_{xy}$ and $J_z$ therefore we cannot associate a global $\mathbb{Z}_2$ symmetry in that case.
But, for the particular situations $\Phi_{i,i+1}=0$ and $\Phi_{i,i+1} = \pi$, ${\cal G}=\prod_i {\cal G}_i$ can yet define a global $\mathbb{Z}_2$ symmetry operator commuting with the Hamiltonian. Consequently, in the absence of a magnetic flux, the sub-lattice symmetry ${\cal R}=\prod_i {\cal G}_i$ becomes a global $\mathbb{Z}_2$ symmetry.

The local current operator $j$ of $a$-particles can be evaluated as the time derivative of the particle densities which are related to $\sigma_i^z$, as we defined $\sigma^z_i = a_{1i}^\dag a_{1i} - a_{2i}^\dag a_{2i} = n^a_{1i}-n^a_{2i}$. This can be computed by
$j = -i[H,\sigma_i^z]$ and gives a parallel component $j_\parallel$ (proportional to $J_{xy}$) and a perpendicular component $j_\perp$ (proportional to $g$).
We assume (if not stated otherwise) in the formulas that for all rungs $i$ we have the magnetic vector potential such that $a'A_{\perp i} = aA^\parallel_{i,i+1}$.
At each site $i$, the perpendicular current and the outgoing parallel current are determined by the following operators \cite{petrescu2013bosonic}
\begin{subequations}\label{first_currents}
\begin{align}
  j_\perp =& -2 g (\sigma_i^x \sin(a'A_{\perp i}) + \sigma_i^y \cos(a'A_{\perp i}) )\label{first_perp_current}, \\
  j_\parallel =& -4i{J}_{xy}(\mathrm{e}^{iaA^\parallel_{i,i+1}} \sigma_i^+\sigma_{i+1}^- - \mathrm{e}^{-iaA^\parallel_{i,i+1}} \sigma_i^-\sigma_{i+1}^+), \nonumber \\
  =&  2{J_{xy}}\bigg( (\sigma_i^x \sigma_{i+1}^x+\sigma_i^y\sigma_{i+1}^y)\sin(aA^\parallel_{i,i+1})\nonumber\\
  &+ (\sigma_{i}^y \sigma_{i+1}^x -\sigma_i^x\sigma_{i+1}^y)\cos(aA^\parallel_{i,i+1})\bigg).\label{first_par_current}
\end{align}
\end{subequations}
Under the application of the $1\leftrightarrow 2$ symmetry, we also verify that the current operators are modified as $j_{\perp}\rightarrow -j_{\perp}$ and $j_\parallel \rightarrow -j_\parallel$.

\par In the following, we investigate how the parallel current behaves under different configurations of the parameters of this model. We regard the parallel current for different strengths in the disorder potential $U_{af}$. In this way, we use it as an indicator for localization in the different regimes. This requires to evaluate the expectation value of the current operator in Eq. $\eqref{first_par_current}$ in the ground state of the system by invoking different approximations, which will be justified below. Finally, we compare also to other indicators.

\section{Weakly-Coupled Rungs Limit}\label{Meissner_effect}
\label{runglimit}

\subsection{Meissner Effect with Static Impurities}\label{static_impurities}
\label{static}

The bosonic ladder model introduced in Eq. \eqref{petrescu_hamilton} in its superfluid phase shows an analogue of the Meissner effect \cite{meissner1933neuer} through the formation of currents along the legs of the ladder proportional to the negative applied flux and a screening of the currents along the rungs \cite{orignac2001meissner, petrescu2013bosonic}.
The existence of a `spin-Meissner'-like phase in the Mott-insulating regime can be understood through the Schwinger-boson representation used to map the model to Eq. \eqref{effective_spin}.
In this framework, the magnetic field couples to the spin degrees of freedom \cite{petrescu2013bosonic}. Lattice models with this property can exhibit a spin current even when the charge sector is in an insulating phase \cite{altman2003phase}. A particle current along one leg is associated to a hole-like current along the other leg. Due to the form of the Josephson term $g$, to minimize energy, the perpendicular current has a zero net-transfer of charge so that $\langle j_{\perp}\rangle=0$. The parallel current screening the $\mathbb{U(1)}$ magnetic flux gives rise to a current of purely spin origin related to the fact that we fix the boson density to unity on a rung  \cite{petrescu2013bosonic}.
Here we consider a setup with small interactions between $a$-particles along the legs, which translates in the language of Hamiltonian \eqref{effective_spin} to $g \gg J_{xy}, J_z$. We thus call this limit the \textit{`weakly-coupled rungs limit'} or \textit{`(almost) decoupled rungs limit'}.\par
Without impurities, the expectation value of the parallel current operator or simply the parallel current can be evaluated invoking a pinning of the phase due to the dominant $g$-term as
\cite{petrescu2013bosonic}
\begin{equation}
	\langle j_\parallel \rangle = -2 J_{xy}\Phi_{i,i+1}.
\end{equation}
In the following we will describe how this result is modified in the presence of an interaction with impurities.
We will first study the case of static impurities to show that in the weakly-coupled rungs limit, the localization shows a smooth power-law profile as a function of $U_{af}$, reflecting the classical aspect of the impurities in this situation. General information on the numerical approach can be found in Appendix A. Then, as a first setup with mobile quantum impurities, we will study the case of vertical motion described by a transverse field for the $\vec{\tau}$-spins. This situation precisely refers to the bouncing between top and bottom legs on a given rung such that the effect of impurities can be solved one by one.\par

\begin{figure}
  \centering
  \includegraphics[width=\linewidth]{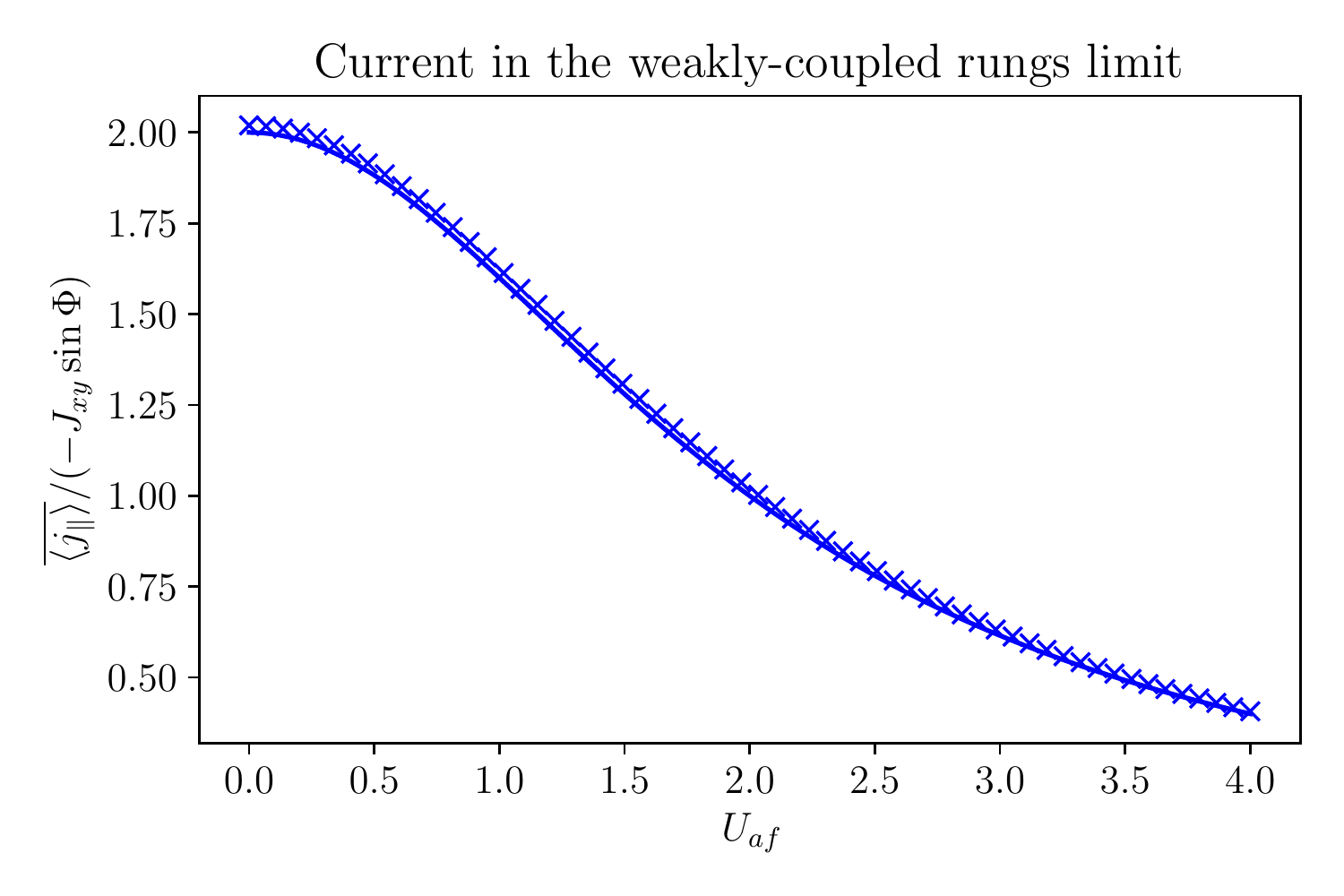}
  \captionof{figure}{The parallel current $\langle j_\parallel \rangle$ as a function of the coupling term $U_{af}$ normalized by $-\sin\Phi$ and $J_{xy}$. Comparison of Eq. \eqref{current_prediction} (solid line) with simulation results (crosses) obtained as an average over all possible configurations of $\tau_i^z$ (we wrote $\overline{\langle j_\parallel\rangle}$ in  the y-axis to signify that here we are considering a disorder average of the expectation value of the current).
  The parameters used for this simulation were $J_{xy}=0.01$, $J_z = 0.01$, $g=1.0$, $a'A_{\perp i} = a A^\parallel_{i,i+1} = 0.01$ for all sites $i$ leading to $\Phi = \Phi_{i,i+1}= -a A^\parallel_{i,i+1} = -0.01$. The simulation was done for a chain with eight sites and periodic boundary conditions. The current was measured between two neighbouring sites $i$, $i+1$.}
	\label{theory_vs_simulation}
\end{figure}

Including the effect of impurities and considering $g$ and $U_{af}$ as a magnetic field in transverse and longitudinal direction respectively, the limit is attained if this magnetic field is large compared to $J_{xy}$ and $J_z$. It is then justified to consider the ground state as the state minimizing the energy on each rung $i$ neglecting the influence of $J_{xy}$ and $J_z$. We can thus consider a Bloch sphere representation of the decoupled spins, which we define by
\begin{subequations}\label{bloch_sphere}
	\begin{align}
		\langle \sigma_i^x \rangle &= \cos\Theta_i \cos\rho_i,\\
		\langle \sigma_i^y \rangle &= \cos\Theta_i \sin\rho_i,\\
		\langle \sigma_i^z \rangle &= \sin\Theta_i.
	\end{align}
\end{subequations}
Here, $\Theta_i$ represents the inclination angle and $\rho_i$ the azimuthal angle with $\Theta_i \in [-\frac{\pi}{2}, \frac{\pi}{2}]$ and $\rho_i \in [0, 2\pi)$. This choice of coordinates is useful to calculate the minimization of the energy, as explained in Appendix \ref{bloch_sphere_calculation}. Requiring minimization of energy yields
\begin{subequations}\label{bloch_sphere_expressions}
\begin{align}
  \rho_i &= -a' A_{\perp i},\\
  \sin \Theta_i &= -\tau_i^z\frac{U_{af}/(2g)}{\sqrt{1+(\tau_i^z U_{af}/(2g))^2}},\\
  \cos \Theta_i &= \frac{1}{\sqrt{1+(\tau_i^z U_{af}/(2g))^2}}.
\end{align}
\end{subequations}
The details of this calculation can be found in appendix \ref{bloch_sphere_calculation}.
We recall that $\tau_i^z = \pm 1$ is just a number and in particular $(\tau_i^z)^2=1$ to simplify further. Plugging the obtained results into Eq. \eqref{first_par_current} and invoking a mean-field approximation for the two-body terms in the current, we finally obtain
\begin{equation}\label{current_prediction}
  \langle j_\parallel \rangle= -2 J_{xy} \frac{1}{1+(U_{af}/2g)^2}\sin\Phi_{i,i+1}.
\end{equation}
Below, we fix the perpendicular component of the vector potential such that for all rungs $a' A_{\perp i} = a' A_{\perp}$ and the parallel component such that $a A^\parallel_{i,i+1} = a A^\parallel$. The flux is similar on all the sites with $\Phi_{i,i+1} = \Phi = -a A^\parallel$ corresponding to a uniform magnetic field in $z$-direction and inducing a homogeneous current in the system. We use such a setup to show the validity of Eq. \eqref{current_prediction} by comparing to results from ED simulations, which is shown in Fig. \ref{theory_vs_simulation}.

Since the decoupled rung limit has been evoked for the calculations leading to Eq. \eqref{current_prediction}, there is consequently no dependence of the current on a site on the respective $\tau_i^z$-variable. Mathematically, this can be understood since the $U_{af}$-term gets a second contribution of $\tau_i^z$ from the approximate solution for $\langle \sigma_i^z \rangle$, so that we finally have $(\tau_i^z)^2 = 1$ as we are treating these variables as static. In the ED simulations, there is still an influence on the configuration of $\tau_i^z$, since they are performed for small, but non-zero values of $J_z$ and $J_{xy}$, which is not addressed in the decoupled rung approximation. For that reason, we compare the prediction to the average over all possible configurations of $\tau_i^z$. Throughout this article if not mentioned differently, we indicate expectation values or (averaged values) by $\langle \cdot \rangle$, while we denote disorder averages of a quantity $A$ over realizations of $\tau_i^z$ disorder configurations by a bar, i.e. $\overline{A}$. \par
In Fig. \ref{theory_vs_simulation}, we show $\overline{\langle j_\parallel \rangle}$ between two neighbouring sites when averaging over all possible configurations of disorder and using periodic boundary conditions. The result \eqref{current_prediction} shows that the current in this {\it semiclassical}-impurity regime reveals a power-law profile, even for large values of $U_{af}$ in agreement with the numerical results in Fig. \ref{theory_vs_simulation}.\par

It is also relevant here to distinguish the present situation from the case of one impurity localized at a given site. In the rung-Mott phase the term $U_{af}$ would produce a renormalization of the on-site energy which can be re-absorbed in the chemical potential $\mu$.

\subsection{Mobile Impurities}
\label{mobile}

In the limit of decoupled rungs the $f$-particles can be easily rendered as quantum particles constrained to hopping along a rung, which technically corresponds to addition of a term $-g_f \tau_i^x$ to the Hamiltonian \eqref{effective_spin}.

Associated to the Ising symmetry $\sigma_i^z\rightarrow -\sigma_i^z$ and $\tau_i^z\rightarrow -\tau_i^z$ when inverting the two legs of the ladder, now we also have $\tau_i^y \rightarrow -\tau_i^y$ such that the quantum algebra for the $\vec{\tau}$ spin is maintained. The operator ${\cal R}$ associated to the sub-lattice symmetry $1\leftrightarrow 2$ and introduced in Eq. (\ref{R}) now takes the form ${\cal R}=\prod_i \sigma_i^x{\cal O}\otimes \tau_i^x$. In addition, the local symmetry of Eq. (\ref{G}) acting on each rung becomes
\begin{equation}
  \mathcal{G}_i = \left(\cos(a'A_{\perp i})\sigma_i^x - \sin(a'A_{\perp i})\sigma_i^y \right)\otimes \tau_i^x,
\end{equation}
commuting with the Hamiltonian when $J_{xy}=J_z=0$. Below, we discuss the implications of this local symmetry for $\mathbb{Z}_2$ lattice gauge theory (LGT), from decoupled rungs.

For each of the decoupled rungs, the situation is then comparable to a model for $\mathbb{Z}_2$ LGT in a double well \cite{barbiero2018coupling, schweizer2019minimal}: Let us define a new spin variable with $\gamma_i^x = \sigma_i^z$ and $\gamma_i^z = \cos(a'A_{\perp i})\sigma_i^x - \sin(a'A_{\perp i})\sigma_i^y$ (which effectively corresponds to a rotation). Then the local symmetry in the decoupled rung limit reads $\mathcal{G}_i = \gamma_i^z \otimes \tau_i^x$.
Each rung then corresponds to a $\mathbb{Z}_2$ LGT on a double well with $\tau_i^z$ as a $\mathbb{Z}_2$ gauge field and $\tau_i^x$ playing the role of a $\mathbb{Z}_2$ electric field \cite{barbiero2018coupling, schweizer2019minimal}. From the Hamiltonian of one rung
\begin{equation}
  H = -g \gamma_i^z -g_f \tau_i^x + U_{af} \gamma_i^x \tau_i^z,
\end{equation}
we see that for $U_{af}=0$, we would have the $\vec{\gamma}$-spin oriented in $z$-direction and the $\vec{\tau}$-spin oriented in $x$-direction, which would lead to an eigenvalue $+1$ of the operator $\mathcal{G}_i$. With non-zero $U_{af}$ the dynamics of both spins are coupled, but since $\mathcal{G}_i$ commutes with the Hamiltonian, its eigenvalue is conserved, therefore flipping $\gamma_i^z$ requires flipping $\tau_i^x$ as well, which comes with an energy cost proportional to $g_f$ \cite{schweizer2019floquet}. This shows how a large parameter $g_f$ can through the coupling of both spins effectively stabilize the $\vec{\gamma}$-spin in the $z$-direction. Transforming back to the $\vec{\sigma}$-variables, this implies that a large $g_f$ term supports the orientation of the $\vec{\sigma}$-spin in the direction given by the $g$ term. Therefore, $g_f$ stabilizes the superfluid spin current and we thus expect it to hinder the localization. \par
To complete the analogy with \cite{schweizer2019floquet, schweizer2019minimal, barbiero2018coupling}, we can define a charge for each site of the double well (which correspond to the different legs) from the operator $\gamma_i^z$ by expressing through a boson tunnelling between the two sites with $\gamma_i^z = \tilde{n}_{1,i} - \tilde{n}_{2,i}$.
Here $\tilde{n}_{\alpha,i}$ is the number operator on the respective site of the rung $i$ and the charge would be defined by $Q_{\alpha,i} = (-1)^\alpha \gamma_i^z$.
We could then define the two conserved local symmetry operators $\mathcal{G}_{i,\alpha} = Q_{\alpha,i} \otimes \tau_i^x$ as in \cite{schweizer2019minimal}. However, since for the situation of one particle in a double well $Q_{1,i}$ and $Q_{2,i}$ are related by $Q_{1,i} Q_{2,i} = -1$ (similarly to \cite{barbiero2018coupling}), we drew conclusions about the influence of this $\mathbb{Z}_2$ LGT on a double well directly using the symmetry operator $\mathcal{G}_i$.
\par It is important to emphasize that the previous considerations only hold for decoupled rungs. When $J_{xy},J_z \neq 0$, our model cannot be described by $\mathbb{Z}_2$ LGT as the operators $\mathcal{G}_i$ do not commute with the Hamiltonian in that case. For the special cases of $\Phi_{i,i+1} = 0$ and $\Phi_{i,i+1} = \pi$, $\mathcal{G} = \prod_i \mathcal{G}_i$ realizes a global $\mathbb{Z}_2$ symmetry which is broken for general values of the flux. \par
In the realm of $\mathbb{Z}_2$ LGT, the vison operator or magnetic field operator is usually defined by $B_p = \prod_{l\in\partial p}\tau_l^z$ (with $\partial p$ referring here to the closed path on a given unit cell) \cite{barbiero2018coupling}.
In our case, we identified a minimal $\mathbb{Z}_2$ LGT for each of the decoupled rungs representing a double well. In general, for one-dimensional $\mathbb{Z}_2$ LGT, the magnetic plaquette term can not be defined \cite{schweizer2019minimal}.
However, if we consider the ladder as a whole, we can define the operator $B_p$ phenomenologically in the same form as above and use it to describe the situation for the f-particles.\par
For the simple case of decoupled rungs, we can evaluate observables analytically.
The Hamiltonian can be diagonalized on each rung and the ground state can be written down, from which expectation values can be evaluated directly.
For the limit of decoupled rungs, we verify from Appendix \ref{mobile_impurities_calculation} that when $g_f\neq 0$ we have $\langle \tau^z_i\rangle=0$ and therefore $\langle \tau^z_i \tau^z_{i+1}\rangle = \langle \tau^z_i \otimes \tau^z_{i+1} \rangle=0$.
This implies that $B_p$ is disordered and has a zero expectation value on a square unit cell, the state corresponds to a vison condensate \cite{barbiero2018coupling}. For $g_f = 0$, in the decoupled rung limit we get a degenerate ground state with $\langle \tau_i^z \rangle = \pm 1$ (see Appendix \ref{mobile_impurities_calculation}), so we get a static configuration of $B_p$. Turning on a small positive value of $J_z$, an anti-ferromagnetic configuration of the $\vec{\sigma}$-spins is favored which is through the $U_{af}$-coupling transmitted to the $\vec{\tau}$-spins. Therefore, in this case we get a static configuration with $B_p = -1$. \par
\begin{figure}[t]
  \centering
  \includegraphics[width=\linewidth]{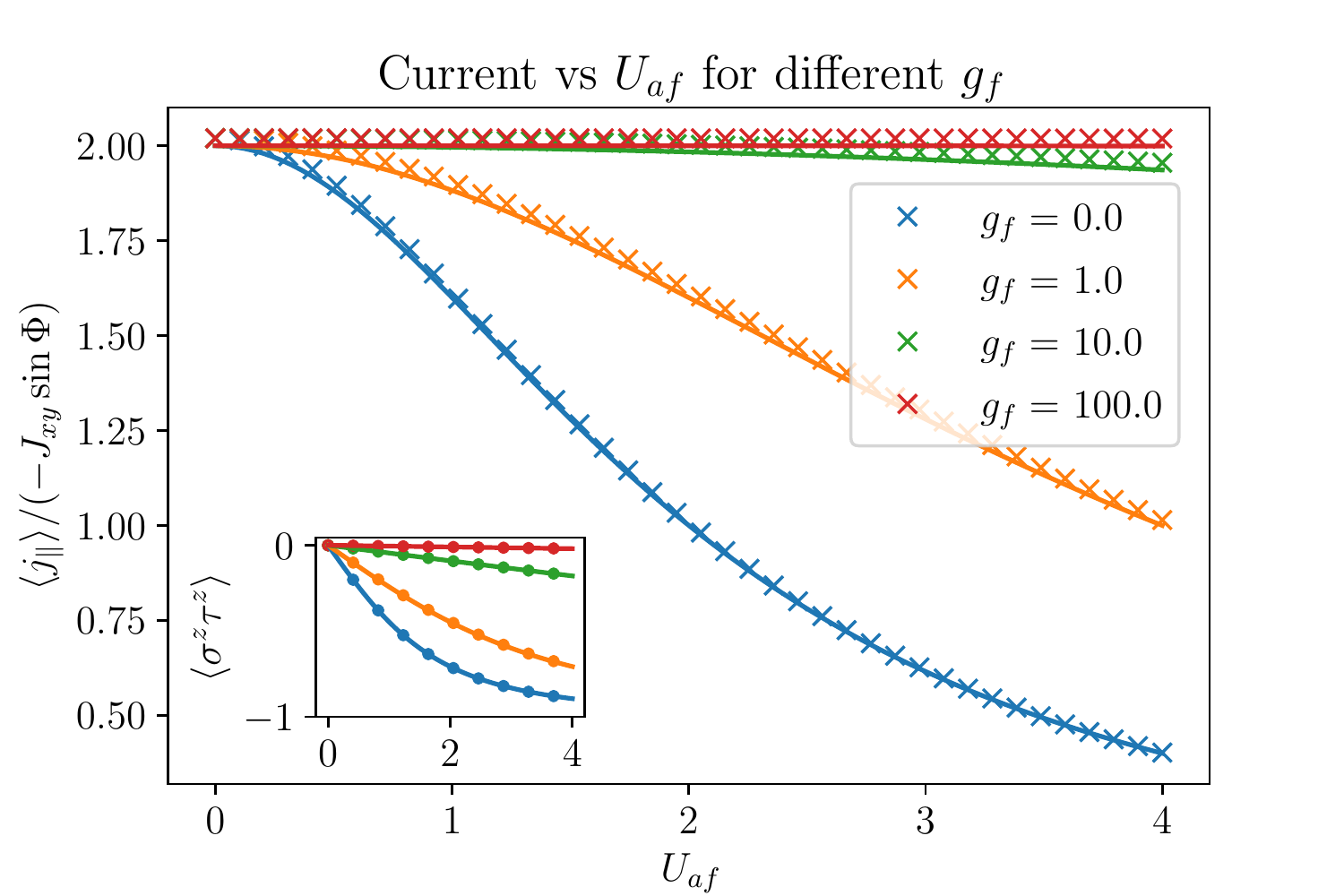}
\caption{Simulation results (crosses) and predictions from Eq. \eqref{par_current_include_gf} (solid lines) for the parallel current as a function of $U_{af}$ for different values of $g_f$ in the ground state of the combined system of $\vec{\sigma}$- and $\vec{\tau}$-spins.
The inset shows the expectation value of the $\sigma_i^z \tau_i^z$-correlation as a function of $U_{af}$, evaluated numerically from ED (dots) and from the analytical result in the decoupled rung limit of Eq. \eqref{correlation_expectation}, for the same values of $g_f$.
Here we simulated a chain with six sites and open boundary conditions and otherwise the same parameters as in Fig. \ref{theory_vs_simulation}. The current and the correlation were evaluated at the center of the chain.}
\label{different_gf}
\end{figure}

We can also introduce a local magnetic quantity $\sigma_i^z \tau^z_i$ which reveals the entanglement of the impurities with the $a$ particles such that $\langle \tau^x_i\rangle <1$. In Fig.  \ref{different_gf}, we study its behavior including small (but finite) values of $(J_{xy},J_z)$ and including a small magnetic field. We compare the numerical findings with analytical results of Appendix \ref{mobile_impurities_calculation} where we detail the calculation of $\langle \sigma_i^z \tau^z_i \rangle$ related to Fig. \ref{different_gf}. We verify that for the impurities we obtain $\langle \sigma_i^z \tau^z_i \rangle^2 + \langle \tau^x_i\rangle^2 =1$ with $\langle \tau^y_i\rangle=0$ since the Hamiltonian is invariant under the transformation $\tau^y_i \rightarrow -\tau^y_i$. Here, from the definition of the Hilbert space with two spins-$\frac{1}{2}$, we have $\langle \tau^\alpha_i\rangle=\langle {\cal I}\otimes \tau^{\alpha}_i \rangle$ with $\alpha=x,y,z$ where
${\cal I}$ is the identity operator or identity $2\times 2$ matrix acting on the Hilbert space of a $\vec{\sigma}$-spin.

Invoking a mean-field approximation for the correlations necessary to compute the parallel current in Eq. \eqref{first_par_current}, we obtain
\begin{equation}
\langle j_\parallel \rangle = -2 J_{xy}\frac{1}{1+\left(\frac{U_{af}/2}{g+g_f}\right)^2} \sin\Phi_{i,i+1}.\label{par_current_include_gf}
\end{equation}
This result can again be verified by comparing to the results from ED simulations which is shown in Fig. \ref{different_gf}. As the impurities here are dynamic quantum objects, they do not play the role of a static disorder but rather enter in the evaluation of the ground state through an extension of the Hilbert space of the spin system. Consequently, Fig. \ref{different_gf} does not show an average over disorder configurations, but the expectation value of the parallel current operator in the ground state. For $g_f=0$, we also verify through Figs. (\ref{theory_vs_simulation}) and (\ref{different_gf}) that the current (density) is identical for periodic and open boundary conditions.

The form of Eq. \eqref{par_current_include_gf} suggests that the current localizes in a similar fashion as for a static $f$-particle configuration, but a large value of $g_f$ protects the current against the effects of strong coupling between $a$- and $f$-particles. This confirms our previous conclusion that due to the conservation of $\mathcal{G}_i$, the $g_f$-term hinders the localization of the current.
In any case, the current in Eq. \eqref{par_current_include_gf} still follows a power-law profile. Note that at $U_{af} = 0$, all curves for the current in Fig. \ref{different_gf} show a small difference between the ED result and the theoretical prediction. This is due to the fact that in the derivation of Eq. \eqref{par_current_include_gf} we considered the rungs completely decoupled for the evaluation of the ground state. In the simulations, we included $J_{xy}$ and $J_z$ with small values. When increasing $U_{af}$, this difference decreases. This can be ascribed to the fact that $U_{af}$ and $g$ act on the spin like an external field whose magnitude increases with increasing $U_{af}$, therefore reducing the influence of $J_{xy}$ and $J_z$.
As described above, a large parameter $g_f$ protects the $\sigma^z$-spin current from the localization induced by $U_{af}$ as it tends to align the $\vec{\tau}$-spins in $x$-direction, which in turn reduces the influence of the $U_{af}$-term, as seen from the inset in Fig. \ref{different_gf}. However, the dynamics of the $\vec{\sigma}$-spins is still determined by the competition between $g$, $J_z$ and $J_{xy}$. This explains why for large values of $g_f$ in Fig. \ref{different_gf} the deviation of the current between ED and theoretical results persists for increasing values of $U_{af}$.\par
In summary, we have seen that in the decoupled rung limit, which is achieved for weak coupling along the legs of the ladder localization occurs on each rung due to a strong interaction of the $a$-particle with the $f$-particle in the form of a power-law, which corresponds to a one-rung localization. \par
Hereafter, we will now keep the quantum property of the impurities, and in the derivation of an effective spin model from the bosonic model in Eq. \eqref{petrescu_hamilton} turn to the limit where $U_{af}$ is not small compared to $U_{aa}$ and $V_\perp$, but of the same order referring to strong interactions with the impurities.

\begin{figure}[t]
\begin{center}
  \includegraphics[width=1\linewidth]{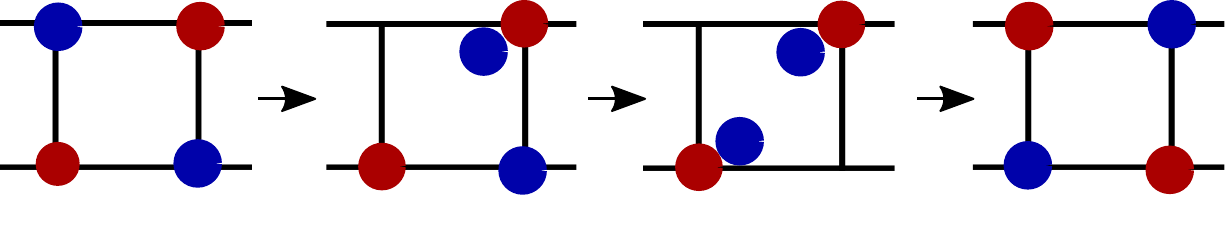}
  \captionof{figure}{A fourth-order correlated hopping of $a$- and $f$-particles, shown with the different colors blue and red respectively, leading to an effective parallel current.}
	\label{fourth_order_exchange}
\end{center}
\end{figure}

\subsection{Strong Interactions with Impurities}\label{large_impurity_interaction}

Here, we will show that when increasing further the interaction strength with impurities, as long as we address the weakly-coupled rungs limit then the sinusoidal response with the $\mathbb{U}(1)$ gauge field remains with a power-law form of the prefactor. In this strong-interaction limit, we derive a four-body spin model showing that the current can keep a similar power-law form. We also describe the effect of the $1\leftrightarrow 2$ symmetry for this situation.

When $U_{af}$ is of the same order as $U_{aa}$ and $V_\perp$ in Eq. \eqref{petrescu_hamilton} and we allow for hopping of the $f$-particles in all directions (and not only along the rungs), we have to account for this when doing the perturbation theory.
If we consider the Hamiltonian without any hopping and at half filling, the ground state is on each rung two-fold degenerate with one $a$- and one $f$-particle on different sites of each rung.

Reintroducing the hopping of $a$-particles perturbatively  again produces second-order Ising interactions such as $J_z^a \sigma_i^z \sigma_{i+1}^z$ with $J_z^a$ different from $J_z$ in the previous sections.
Now, introducing a hopping for the impurities (spinless fermions) along both legs and rungs with $- t_x^f \sum_{\alpha,i} f^\dag_{\alpha i} f_{\alpha,i+1 }- t^f_y \sum_{i} f^{\dag}_{2 i} f_{1 i} +\text{h.c.}$ with $\alpha=1,2$, we get a similar term
\begin{equation}
 J_z^f \tau_i^z \tau_{i+1}^z.
\end{equation}
The hopping term $t_y^f$ can be identified with the parameter $g_f$ in the spin language of Sec. \ref{mobile}.

The interchange of an $a$- and an $f$-particle along a rung is accounted for by a term
\begin{equation}
-g^{af} \mathrm{e}^{ia'A_{\perp i}} \sigma_i^+ \tau_i^- + \text{h.c.},
\end{equation}
where $g^{af}$ is defined differently than $g$ in the previous sections. Interestingly, the correlated limit with large $U_{af}$ produces a 4-particles correlated hopping term to respect all the interaction terms, as shown exemplarily in Fig. \ref{fourth_order_exchange}.
This gives a contribution which reads $-J_{xy}^\parallel \mathrm{e}^{ia(A^1_{i,i+1}-A^2_{i,i+1})}\sigma_i^+\tau_i^-\sigma_{i+1}^-\tau_{i+1}^+ + \text{h.c.}$.\par
\begin{figure}[t]
  \includegraphics[width=\linewidth]{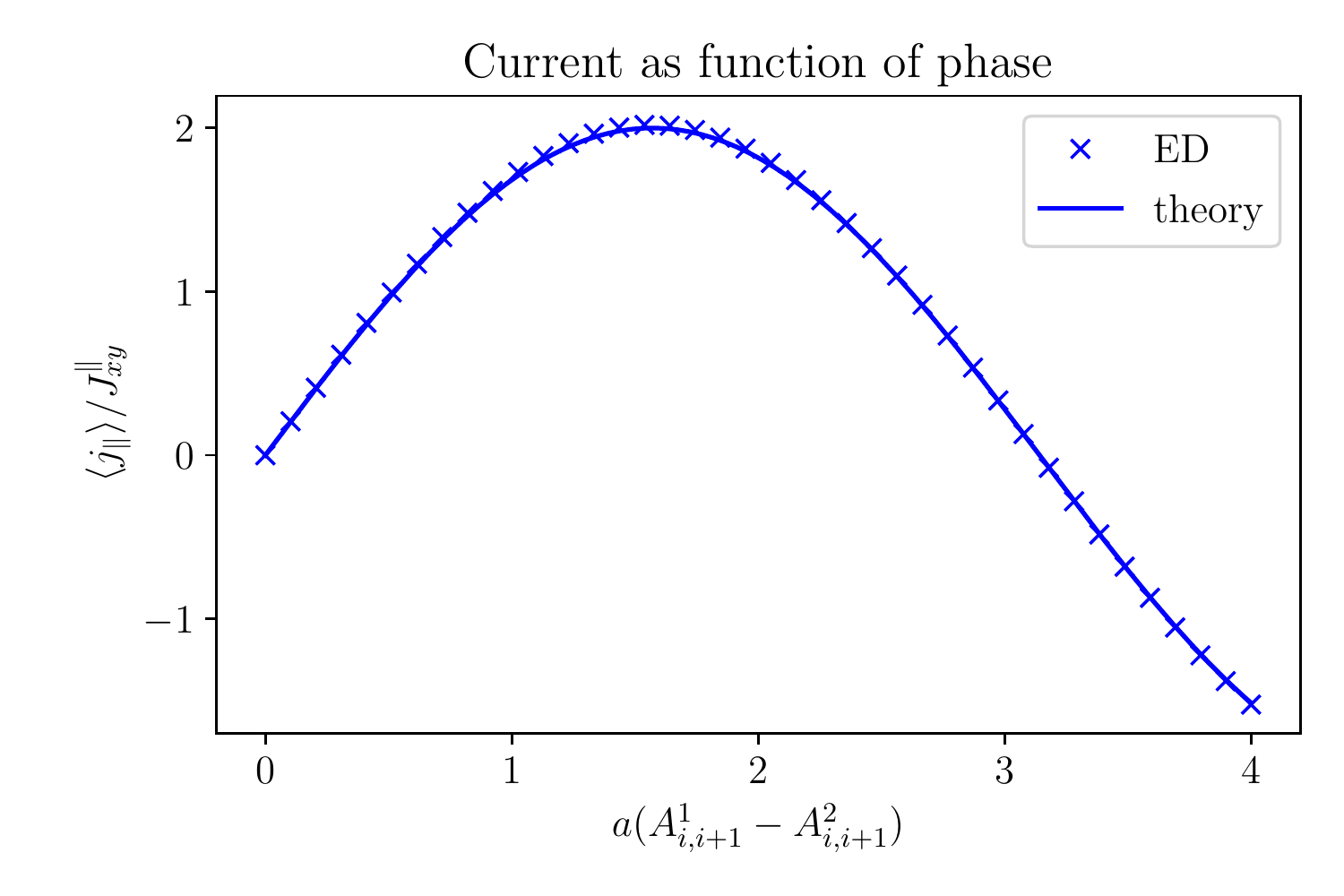}
  \caption{Meissner current in the limit of large $U_{af}$ with the theory from Eq. \eqref{fourbody_current} compared to ED data (crosses) in the ground state of the combined system of $\vec{\sigma}$- and $\vec{\tau}$-spins. The ED data was obtained for a chain with six sites with periodic boundary conditions.
  The chosen parameters were $J_{xy}^\parallel = J_z^a = J_z^f = 0.01$, $g^{af} = 1.0$ and $a'A_{\perp,i} = a(A^1_{i,i+1}-A^2_{i,i+1})$ for all $i$ so that $\Phi_{i,i+1} =\Phi= -a(A^1_{i,i+1}-A^2_{i,i+1})$.}
	\label{current}
\end{figure}

The expressions for the new parameters $J_z^a$, $J_z^f$, $g^{af}$ and $J_{xy}^\parallel$ and their derivation can be found in Appendix \ref{fourbody_derivation} for the sake of clarity. We write the effective Hamiltonian as
\begin{widetext}
\begin{equation}\label{effective_four_body}
  H = -J_{xy}^\parallel \sum_i \mathrm{e}^{ia(A^1_{i,i+1}-A^2_{i,i+1})}\sigma_i^+\tau_i^-\sigma_{i+1}^-\tau_{i+1}^+ + \text{h.c.} + J_z^a \sum_i \sigma_i^z \sigma_{i+1}^z + J_z^f \sum_i \tau_i^z \tau_{i+1}^z -g^{af}  \sum_i \mathrm{e}^{ia'A_{\perp i}} \sigma_i^+ \tau_i^- + \text{h.c.}.
\end{equation}
The parallel current operator can be evaluated from the Hamiltonian as:
\begin{multline}
  j_\parallel = -\frac{i{J}_{xy}^\parallel}{4} (\mathrm{e}^{ia(A^1_{i,i+1}-A^2_{i,i+1})} (\sigma_i^x\tau_i^x +\sigma_i^y\tau_i^y -i\sigma_i^x\tau_i^y + i \sigma_i^y\tau_i^x)(\sigma_{i+1}^x \tau_{i+1}^x +\sigma_{i+1}^y \tau_{i+1}^y -i\sigma_{i+1}^y \tau_{i+1}^x +i\sigma_{i+1}^x \tau_{i+1}^y )\\
  - \mathrm{e}^{-ia(A^1_{i,i+1}-A^2_{i,i+1})}(\sigma_i^x\tau_i^x +\sigma_i^y\tau_i^y +i\sigma_i^x\tau_i^y - i \sigma_i^y\tau_i^x)(\sigma_{i+1}^x \tau_{i+1}^x +\sigma_{i+1}^y \tau_{i+1}^y +i\sigma_{i+1}^y \tau_{i+1}^x -i\sigma_{i+1}^x \tau_{i+1}^y )).
  \label{fourbody_current_full}
\end{multline}
\end{widetext}
When $g^{af}$ is the dominant term, in the ground state we have on each rung
\begin{multline}
\label{equality1}
  \frac{\cos (a'A_{\perp i})}{2} \langle \sigma_i^x \tau_i^x + \sigma_i^y \tau_i^y\rangle\\ + \frac{\sin (a'A_{\perp i})}{2} \langle \sigma_i^x \tau_i^y - \sigma_i^y \tau_i^x\rangle = 1.
\end{multline}
This suggests to write
\begin{subequations}
  \label{equality2}
\begin{align}
  \langle \sigma_i^x \tau_i^x + \sigma_i^y \tau_i^y\rangle &= 2\cos (a'A_{\perp i}),\\
  \langle \sigma_i^x \tau_i^y - \sigma_i^y \tau_i^x\rangle &= 2\sin (a'A_{\perp i}),
\end{align}
\end{subequations}
\color{black}
such that the expectation value of the parallel current reads
\begin{equation}
\langle  j_\parallel \rangle= -2 {J}_{xy}^\parallel \sin \Phi_{i,i+1}.
  \label{fourbody_current}
\end{equation}
This form agrees with a perfect superfluid in agreement with ED from Fig. \ref{current}. In this limit, we observe
a renormalization (reduction) of the prefactor $J_{xy}^{\parallel}$ (see Appendix \ref{fourbody_derivation}). The equalities (\ref{equality1}) and (\ref{equality2}) respect the $1\leftrightarrow 2$ symmetry between the two legs (or wires) of the ladder. On the other hand, in the strong-interaction limit, the physics is also similar as if both the $a$-and $f$-particles participate to the screening of the applied $\mathbb{U}(1)$ gauge field through the
$J_{xy}^{\parallel}$ term. More precisely, setting $J_z^a=J_z^f=0$ in Eq. (\ref{effective_four_body}) which can be realized adjusting appropriately the interactions, we can redefine a spin variable so that $\tilde{\sigma}_i^+ = \sigma_i^+\tau_i^-=\sigma_i^+\otimes \tau_i^-$,  $\tilde{\sigma}_i^- = \sigma_i^-\tau_i^+=\sigma_i^-\otimes \tau_i^+$ and $\tilde{\sigma}^z_i = \sigma^z_i\tau^z_i$ such that the $1\leftrightarrow 2$ transformation is identical to $\tilde{\sigma}_i^+ \leftrightarrow \tilde{\sigma}_i^-$ with $\tilde{\sigma}_i^z$ unchanged and $[\tilde{\sigma}_i^+,\tilde{\sigma}_i^-]=\tilde{\sigma}_i^z$ or equivalently $[\tilde{\sigma}_i^-,\tilde{\sigma}_i^+]=-[\tilde{\sigma}_i^+,\tilde{\sigma}_i^-]=-\tilde{\sigma}_i^z$. \par
With this transformation, in that case the Hamiltonian of two different particle species can be reduced to a Hamiltonian of one spin-degree of freedom for each rung. We observe here that even though we cannot differentiate the impurities from the particles (matter) the current yet takes a similar sinusoidal form.\par
For a strong value of $g^{af}$, then the ground state satisfies $\langle \tilde{\sigma}_i^x\rangle = \cos(a' A_{\perp}^i)$ and
$\langle \tilde{\sigma}_i^y\rangle = -\sin(a' A_{\perp}^i)$ such that the system will remain in the $xy$-plane even in the presence of a finite (small) perturbation $J_z^a$ or $J_z^f$. If we develop the partition function to second-order in $(J_z^a J_z^f)$, then we may have corrections to the Hamiltonian proportional to $\tilde{\sigma}_i^z \tilde{\sigma}_{i+1}^z$ similarly as an $XXZ$ spin chain with a $\mathbb{U}(1)$ gauge field and a transverse magnetic field $g^{af}$. The model takes a similar form as the rung-Mott phase Hamiltonian \cite{petrescu2013bosonic} which then gives another interpretation to the current in Eq. (\ref{fourbody_current}).
This analysis then shows that as long as we consider weakly-coupled rungs, the system can be described through local observables as a result of the transverse magnetic field in the spin representation.

\section{Strongly-Coupled Rung Model}
\label{stronginteractions}

Here, we study the regime of strong inter-rung interactions for the situation with a telegraph potential or two-peak disorder. To have strongly-coupled rungs, we adjust $J_{xy} \gg g$ for the $a$ particles in Eq. (\ref{effective_spin}) comparing different distributions (configurations) for the impurities, and for simplicity start with $g = 0$. Here, we can anticipate a persistent current going along the chain with periodic boundary conditions in the ground state due to the magnetic field which entered through the Peierls substitution into Eq. \eqref{effective_spin} \cite{schulz1995fermi}. We study the localization regime for the $a$-particles driven by a large $U_{af}$ interaction for specific configurations of the static impurities. The objective is to compare with the situation of a one-dimensional Gaussian disorder potential \cite{GiamarchiSchulz}. The situation with finite $g$ will be addressed in Sec. \ref{MBL} related to many-body localization.

\subsection{The persistent current limit}

If we assume the $\tau_i^z$-variables to be statically fixed to $\pm 1$ and randomly distributed, the model corresponds to a XXZ-chain with a $\mathbb{U}(1)$ gauge field and random magnetic field in z-direction:
\begin{widetext}
\begin{equation}\label{effective_XXZ}
  H = \sum_i \left(-2J_{xy}\mathrm{e}^{i\phi} \sigma_i^+ \sigma_{i+1}^- +\text{h.c.} + J_z \sigma_i^z \sigma_{i+1}^z + \frac{U_{af}}{2} \sigma_i^z \tau^z_i\right).
\end{equation}
\end{widetext}
Here, we define $aA_{i,i+1}^\parallel=-\Phi=\phi$ on all rungs $i$. The $1\leftrightarrow 2$ transformation defined in Sec. \ref{rungMottDefinitions} is also equivalent to $\phi\leftrightarrow -\phi$ with $\sigma_i^+\leftrightarrow \sigma_i^-$.
This model without the $\mathbb{U}(1)$ gauge field has been studied in relation with many-body localization and remains an active area of research \cite{alet2018many}.
Numerically, the model \eqref{effective_XXZ} has been considered with $\phi = 0$ and $-J_{xy} = J_z = 0.25$ and a random field of the form $h_i \sigma_i^z$, where $h_i$ are drawn from a uniform distribution \cite{pal2010many,luitz2015many, bardarson2012unbounded}.
A transition to a many-body localized phase has been found for strong disorder. The same model was used in a recent study suggesting the use of bipartite entanglement and fluctuations to characterize this transition \cite{singh2016signatures}. Here we study the model with a peaked disorder (characterized by $\tau_i^z = \pm 1$) with the energy scale given by $U_{af}$. We could for example assume that the $\tau_i^z$ are at each site independently drawn Bernoulli variables with a universal or site-dependent probability to give $\pm1$ depending on the physical situation. Considering the $\vec{\sigma}$-spin current between two sites, it varies depending on the configuration of $\vec{\tau}$-spins on these two sites. This is also confirmed by simulations, as shown in Fig. \ref{ED_comparison_ordered}.\par
In order to make progress analytically, now we study two particular cases of configurations, which are in this setup {\it ferromagnetic} and {\it antiferromagnetic} order for the impurities and we will compare the results to those obtained when averaging over various disorder configurations.
This will reveal that localization indeed occurs in a steep manner.
The word steep also refers to an insulating phase for the $a$-particles, with a zero current.

\subsection{Jordan-Wigner Transformation}\label{ferro_order_section}

Assume that on all rungs, we start with a uniform or ferromagnetic $\tau_i^z = \tau^z = \pm1$ situation, so that all $f$-particles live on one leg of the ladder.
For $J_z = 0$, the ground state can be found using a mapping to spin-less fermions due to Jordan and Wigner \cite{lieb1961two}.
The Jordan-Wigner transformation
\begin{eqnarray}
  \sigma_i^+ &\rightarrow& c_i^{\dag} \exp \left(i\pi \sum_{l<i} n_l\right), \nonumber \\
  \sigma_i^z &\rightarrow& 2n_i-1,
\end{eqnarray}
maps the Hamiltonian \eqref{effective_XXZ} to
\begin{equation}\label{JW_transformed}
  H = - 2 J_{xy}\sum_i \mathrm{e}^{i \phi}  c_i^{\dag} c_{i+1} + \text{h.c.} + \frac{U_{af}}{2}\sum_i (2n_i-1) \tau_i^z.
\end{equation}
It can easily be verified that the $c_i$ and $c_i^\dag$ fulfill fermionic anticommutation relations, which is ensured by the string term $\exp \left(i\pi\sum_{l<i} n_l\right)$.
Some care needs to be taken about the boundary conditions of the spin chain model: For open boundary conditions, Eq. \eqref{JW_transformed} holds true, but if we want to incorporate periodic boundary conditions, we need to add a term $\mathrm{e}^{i \phi} \sigma_N^+ \sigma_1^- + \text{h.c.}$ in the spin chain Hamiltonian \eqref{effective_XXZ}. After the Jordan-Wigner mapping, it reads
\begin{multline}
    \mathrm{e}^{i \phi} c_N^\dag \mathrm{e}^{i \pi \sum_{l<N} n_l} c_1 + \mathrm{e}^{-i \phi} c_1^\dag \mathrm{e}^{-i \pi \sum_{l<N} n_l} c_N \\
    = \mathrm{e}^{i(\pi(m-1)+\phi)}  c_N^\dag c_1 + \text{h.c.},
\end{multline}
where $m$ denotes the total number of fermions, which is still constant, but according to its parity, we get periodic or anti-periodic boundary conditions in the free fermion model. We can then write the full Hamiltonian as
\begin{widetext}
\begin{eqnarray}\label{JW_transformed_pbc}
  H = - 2{J_{xy}} \left[ \sum_{i=1}^{N} \mathrm{e}^{i \phi}  c_i^{\dag} c_{i+1} + \text{h.c.}  -(\mathrm{e}^{i\pi m}+1)(\mathrm{e}^{i \phi}  c_N^{\dag} c_{1} + \text{h.c.})\right] + \frac{U_{af}}{2}\sum_i (2n_i-1) \tau_i^z.
\end{eqnarray}
\end{widetext}
The $1\leftrightarrow 2$ transformation now corresponds to $c_i\leftrightarrow c_i^{\dagger}$ with $\phi\rightarrow -\phi$.
We can then go on and diagonalize the Hamiltonian by Fourier transforming separately in the sectors where $m$ is even and odd.
\par

This can be done by introducing Fourier transformed operators:
\begin{equation}
    c_j = \frac{1}{\sqrt{N}} \sum_k \mathrm{e}^{-ikaj}c_k,
\end{equation}
with $k=2\pi n/(Na)$ if $m$ is even and with $k=\frac{2\pi}{a} (\frac{n}{N}-\frac{1}{2})$ if $m$ is odd to account for the anti-periodicity of the boundary conditions, with $n=0,1,...,N-1$. In the following, we assume that $m$ is odd and keep in mind that the calculations can easily be generalized to the case where $m$ is even. The Hamiltonian \eqref{JW_transformed} can then upon performing the summation and neglecting a constant term easily be brought to the form
\begin{equation}\label{Hamiltonian_diagonal}
    H = \sum_k \omega(k) c_k^\dag c_k
\end{equation}
with $\omega = - 4 J_{xy} \cos(ka-\phi) + U_{af}\tau^z$ and $a$ the lattice spacing of the spin chain.

\begin{figure}
  \centering
  \includegraphics[width=0.25\textwidth]{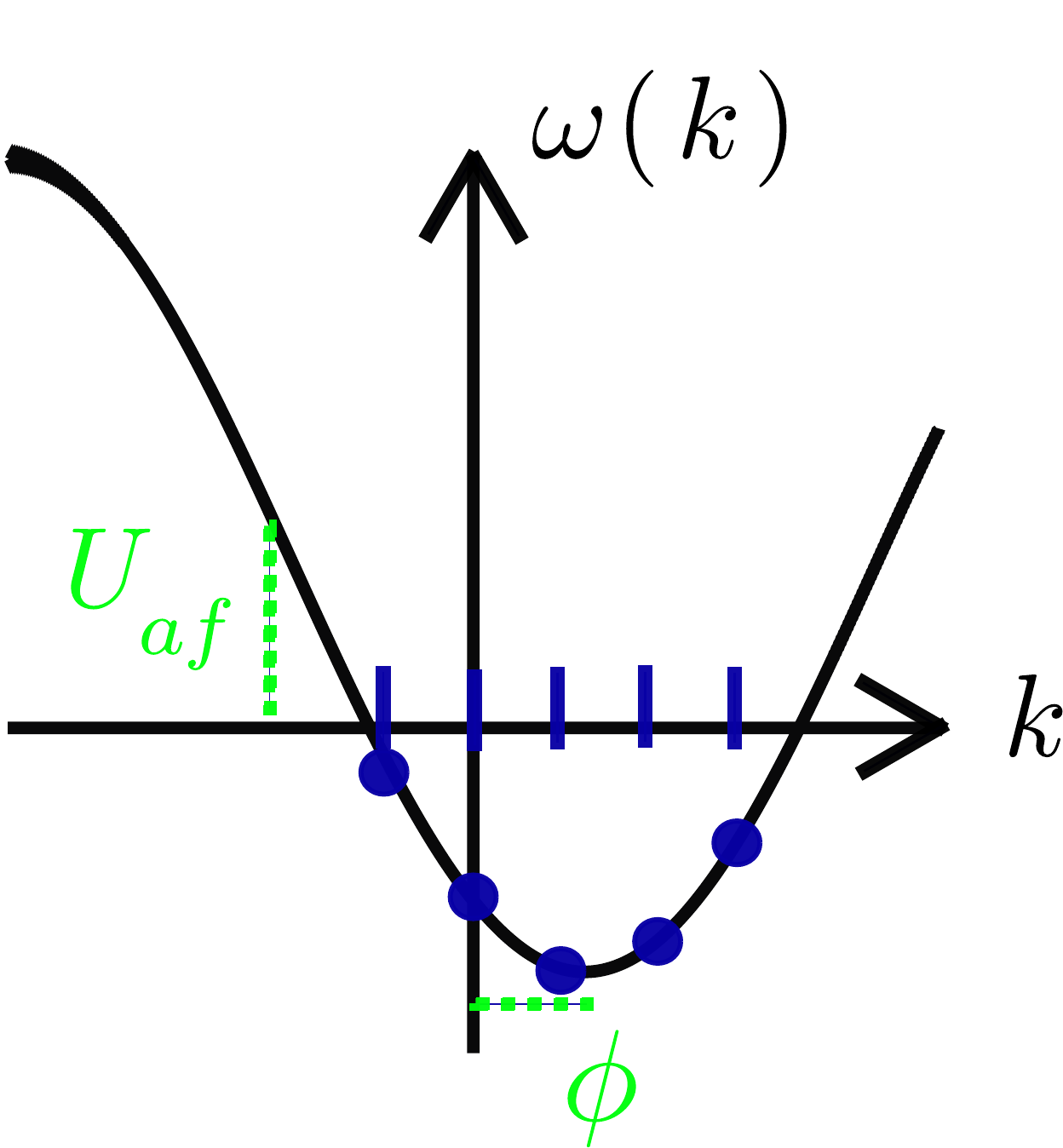}
  \captionof{figure}{The band is shifted by $\phi/a$ in k-direction and by $U_{af}$ in the energy, the latter thus has the effect of a chemical potential. The occupied states are shown in blue. The Fermi momenta change accordingly.}
	\label{band}
\end{figure}

The current calculated from the time derivative takes the form:
\begin{widetext}
\begin{eqnarray}
    j_i = 4 i J_{xy}\left( \mathrm{e}^{i\phi} \sigma_{i}^+ \sigma_{i+1}^- - \mathrm{e}^{-i\phi} \sigma_{i}^- \sigma_{i+1}^+
    - \mathrm{e}^{i\phi} \sigma_{i-1}^+ \sigma_{i}^- + \mathrm{e}^{-i\phi} \sigma_{i-1}^- \sigma_{i}^+ \right).
\end{eqnarray}
\end{widetext}
The spin current in this basis at a site $i$ is calculated by
\begin{eqnarray}\label{true_current}
    I_i &=& -i[H,\sigma_i^z] = -i[H,2n_i -1] \nonumber \\
    &=& 4 i J_{xy} \left(\mathrm{e}^{i\phi} c_{i-1}^\dag c_i - \mathrm{e}^{i\phi}c_{i}^\dag c_{i+1} - \text{h.c.} \right).
\end{eqnarray}
We call the current flowing out from one side $j_i$, therefore $I_i = j_i -j_{i-1}$ and we can conclude from this and Eq. \eqref{true_current} that in the ground state (in equilibrium):
\begin{equation}\label{current_derivative}
j= \frac{2}{N}\pdv{H}{\phi}.
\end{equation}
In the ground state, negative energy states will be occupied, i.e.
\begin{equation*}
     4 J_{xy} \cos(a k-\phi) > U_{af}\tau^z,
\end{equation*}
which is fulfilled for $k$ between $k_\pm = \phi/a \pm \arccos\left( \frac{U_{af} \tau_i^z }{4 J_{xy}} \right)/a$. This is shown in Fig. \ref{band}.

The current reads
\begin{equation}\label{current_sum}
    j = \frac{2}{N}\pdv{H}{\phi} = -4J_{xy} \frac{2}{N} \sum_k \sin{(a k-\phi)}c_k^\dag c_k.
\end{equation}
For large N, we can write the expectation value of the current operator as
\begin{align}\label{momentum_integral}
    \langle j \rangle &=  -\frac{4J_{xy}}{\pi} \int_{k_-}^{k_+} dk \sin{(ak-\phi)} \nonumber \\
    &= \frac{4J_{xy}}{\pi} \cos(ak-\phi)\bigg\rvert_{k_-}^{k_+} = 0 ,
\end{align}
which vanishes in the continuum limit. To explain the current in a finite-size system, we have to stay with the non-continuous case: the allowed momentum values in the first Brillouin zone (for the case where the fermion number $m$ is odd) are $k = 2 \pi n /(Na) - \pi/a$ with $n = 0,1,...,N-1$ (or equivalently $2\pi r /L$ with $r=-N/2, -N/2 + 1,... N/2-1$), so the momenta are spaced with $2 \pi / L$, according to Fig. \ref{band}. In order to get a better approximation to the sum in Eq. \eqref{current_sum}, we can attempt to do the integration as in Eq. \eqref{momentum_integral} exactly between the outermost two occupied states. We therefore have to change the integration boundaries to

\begin{eqnarray}
    \tilde{k}_- &=& k_- - \left(k_-\mod \frac{2\pi}{L} \right), \nonumber \\
    \tilde{k}_+ &=& k_+ - \left(k_+\mod \frac{2\pi}{L}\right).
\end{eqnarray}
We obtain the current
\begin{widetext}
\begin{eqnarray}
    \langle j \rangle \approx \frac{4 J_{xy}}{\pi} \left[ \frac{U_{af} \tau_i^z }{4 J_{xy}}(-2\sin\phi \sin\alpha) + \sqrt{1-\left(\frac{U_{af} \tau_i^z }{4 J_{xy}}\right)^2}(2\sin \phi \cos \alpha) \right],
\end{eqnarray}
\end{widetext}
where $\alpha = \arccos{\left(\frac{U_{af} \tau_i^z }{4 J_{xy}} \right)} \mod \frac{2\pi}{L}$.

Assume $L$ is large, therefore $\alpha$ is small and if furthermore we consider only small phases, the first term becomes negligible and in the second term we can write $\cos\alpha \sim 1$, therefore obtaining
\begin{equation}\label{simple_current_theory}
     \langle j \rangle = 8\frac{J_{xy}}{\pi} \sqrt{1-\left(\frac{U_{af} \tau_i^z }{4 J_{xy}}\right)^2} \phi.
\end{equation}

This result can be compared to the results from ED simulations, which can be seen from the solid orange curve in Fig. \ref{ED_comparison_ordered} which shows Eq. \eqref{simple_current_theory}. The orange crosses show ED results for a setup where on all sites $\tau_i^z = 1$, where we averaged over all sites of the system to make the connection with the calculation in momentum space. Even though in this case, the current has the same sign as in the decoupled rung limit, it is clear by comparing the form of Eq. \eqref{current_prediction} to Eq. \eqref{simple_current_theory} that both regimes are very different, which is also confirmed by simulations. From Fig. \ref{ED_comparison_ordered} (in orange), we see that the simulation results and Eq. \eqref{simple_current_theory} agree for the {\it steep localization} of the current when $U_{af} \sim 4 J_{xy}$, i.e. the current now goes to zero. The step-like behavior in ED represented through orange crosses reflects Fig. \ref{band}.

When taking an average over many different realizations of $\tau_i^z$ (the blue curve in Fig.  \ref{ED_comparison_ordered}), this behaviour is changed, but we see that there is still a {\it strong localization} effect for the same value of $U_{af}$.
It is interesting to observe that the insulating (or localized) regime also occurs in this case.

In order to make more precise statements, we have to consider also other configurations. Precise results can be obtained for the opposite case of alternating $\tau_i^z$-spins, as shown hereafter. We will show
that interaction effects through $U_{af}$ favor a strong localization effect similar as in the situation of a Gaussian disorder, as also shown through the green curve of Fig. \ref{ED_comparison_ordered}).

\subsection{Alternating $\tau_i^z$ variables}

First, we address the situation of an alternating or staggered potential and solving the model in the reciprocal space directly. If we set $J_z = 0$ for now, we have
\begin{equation}
\label{modelantiferro}
    H = -2 J_{xy} \sum_i \mathrm{e}^{i\phi} c_i^{\dagger} c_{i+1} +\text{h.c.} + U_{af} \sum_i (-1)^i n_i,
\end{equation}
\begin{widetext}
after a Fourier transform with $x_j = aj$
\begin{eqnarray}
    H &=& -4J_{xy} \sum_k \cos(ak-\phi) c_k^\dag c_k + U_{af} \sum_j \sum_{k,k'} \mathrm{e}^{i(k - k' + \frac{\pi}{a})x_j}c_k^\dag c_{k'}, \nonumber\\
    &=& -4J_{xy} \sum_k \cos(ak-\phi) c_k^\dag c_k + U_{af} \sum_{k}c_k^\dag c_{k+\frac{\pi}{a}},
\end{eqnarray}
or in a Bogoliubov-de-Gennes form,
\begin{eqnarray}
    H &=  \sum_{k<0} (c_k^\dag \, c_{k+\frac{\pi}{a}}^\dag) \begin{pmatrix}
-4J_{xy}\cos(ak-\phi) & U_{af} \\
U_{af} & -4J_{xy}\cos(ak-\phi+\pi)
\end{pmatrix}
\begin{pmatrix}
c_k \\
c_{k+\frac{\pi}{a}}
\end{pmatrix}.
\end{eqnarray}
\end{widetext}

\begin{figure}[ht]
  \centering
  \includegraphics[width=\linewidth]{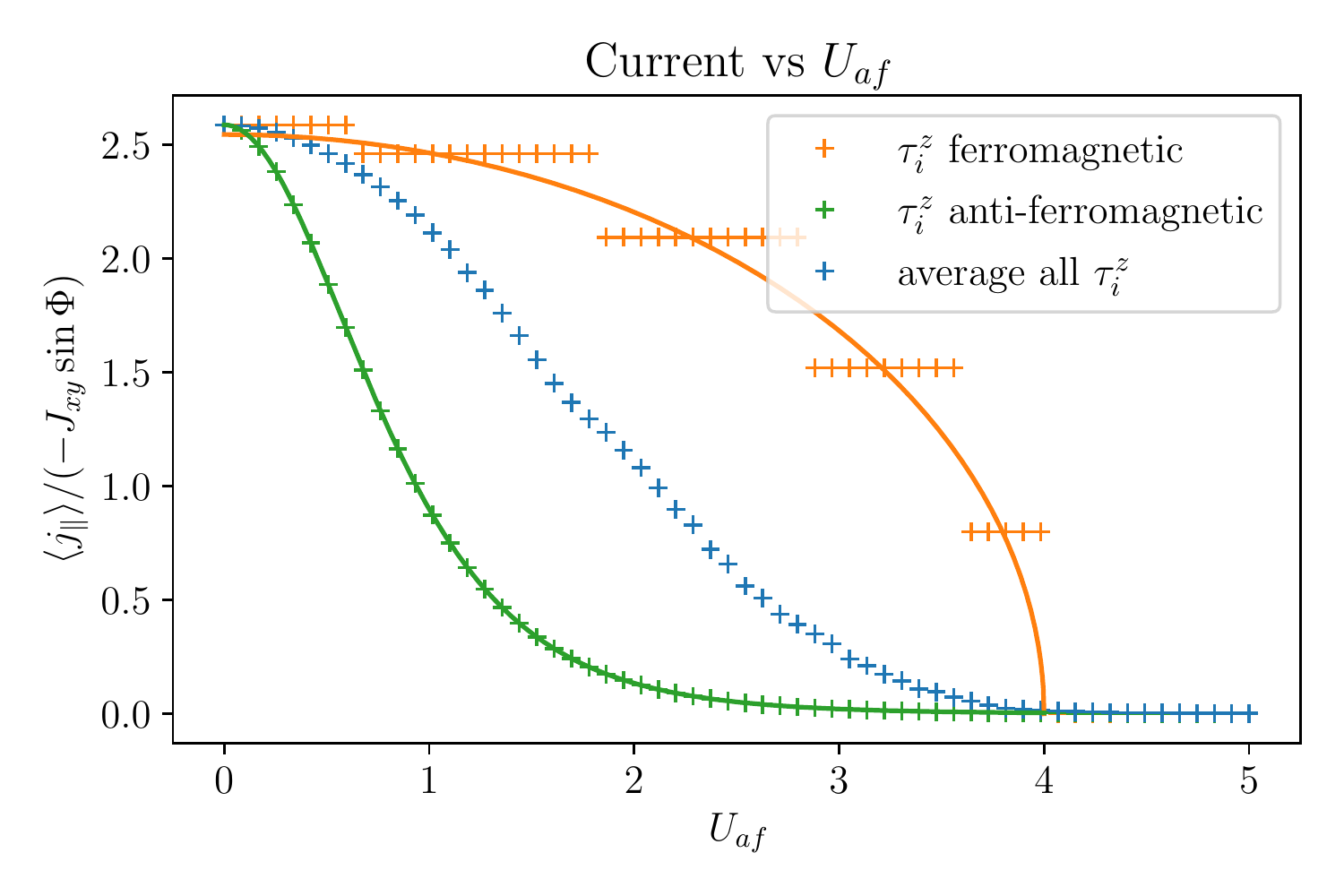}
  \captionof{figure}{Current averaged over all sites of the system from ED as a function of $U_{af}$ for a ferromagnetically (orange crosses) and an antiferromagnetically (green crosses) ordered configurations of $\tau_i^z$-variables. The blue crosses show an average ($\overline{\langle j_\parallel \rangle}$) over all possible configurations of $\tau_i^z$ (here the number of sites $N=10$, so there are $2^{10} = 1024$ possible configurations). We further used $J_{xy} = 1.0$, $J_z = g = 0.0$, $\phi = 0.01$ and periodic boundary conditions. The analytical formula \eqref{simple_current_theory} obtained for a ferromagnetic configuration is the orange solid line, while Eq. \eqref{alternating_current_sum} is shown by the green solid line.}
	\label{ED_comparison_ordered}
\end{figure}

This can easily be diagonalized to obtain the eigenenergies
\begin{equation}
  E_\pm = \pm \sqrt{U_{af}^2 + 16J_{xy}^2 \cos^2(ak-\phi)}.
\end{equation}
In the ground state, all states giving rise to a negative energy contribution will be occupied in the diagonal basis. The ground state energy is thus
\begin{equation}\label{staggered_eigenenergies}
  E = -\sum_{k} \sqrt{U_{af}^2 + 16J_{xy}^2 \cos^2(ak-\phi)}.
\end{equation}
Here the domain of $k$ should have an extension of $\pi/a$.
Due to the symmetry of the squared-cosine function appearing in Eq. \eqref{staggered_eigenenergies}, we can sum over any connected region with extension $\pi/a$ equivalently, in particular we can choose $\phi/a \pm \pi/(2a)$ as summation boundaries which corresponds to those used in Eq. \eqref{current_sum} with $U_{af}=0$.
The current can readily be evaluated as a derivative of equation \eqref{staggered_eigenenergies} with respect to $\phi$ and using Eq. \eqref{true_current}. It reads
\begin{equation}\label{alternating_current_sum}
j = -\frac{2}{N}\sum_{k}\frac{16J_{xy}^2 \cos(ak-\phi)\sin(ak-\phi)}{ \sqrt{U_{af}^2 + 16J_{xy}^2 \cos^2(ak-\phi)}}.
\end{equation}
For $U_{af}=0$, we get back Eq. \eqref{current_sum}.
This result can be compared to the results from ED simulations, which can be seen from the solid green curve in Fig. \ref{ED_comparison_ordered} which shows Eq. \eqref{alternating_current_sum} for $10$ sites.
The green crosses show ED results for an alternating configuration $\ket{\uparrow \downarrow \uparrow \downarrow \uparrow \downarrow \uparrow \downarrow \uparrow \downarrow}$ of the $\tau_i^z$-spins.
To make the connection with the calculation in momentum space leading to Eq. \eqref{alternating_current_sum}, we averaged over all sites of the system.

However, Eq. \eqref{alternating_current_sum} holds only for $J_z = 0.0$. To depart from this special case, we need to resort to different methods, which we will describe in the following.

\subsection{Bosonization}

We now attempt to investigate the special case of disorder treated in the previous Section while including the interaction term proportional to $J_z$. Our goal is to understand how the sum on momenta evolves in the presence of interactions between fermions mediated
by the $J_z$ term here. For this purpose, we will develop a bosonized theory, which simplifies the understanding of these four-body terms. In the next Section, we will then apply Renormalization Group arguments. To develop this framework, start with the interacting fermion-model
\begin{eqnarray}\label{interacting_fermion_hamilton}
    H &=& -2 J_{xy} \sum_i c_i^\dag c_{i+1} +\text{h.c.} + J_z \sum_i (2n_i-1)(2n_{i+1}-1) \nonumber \\
    &+& \frac{U_{af}}{2} \sum_i \tau_i^z (2 n_i-1),
\end{eqnarray}
which resembles the Hamiltonian \eqref{effective_spin} in fermionic language with $g=0$ and without the Peierls phases. For simplicity here, we set the $\mathbb{U}(1)$ gauge field $\phi\rightarrow 0$ and we will comment on the effect of $\phi$ at the end of Sec. \ref{RG}.

As we described in Sec. \ref{ferro_order_section}, the Jordan-Wigner transformation maps between spin operators and fermionic operators on a chain. We can identify the spin raising and lowering operators with creation and annihilation fermion operators
\begin{equation}\label{fermion_of_boson}
    c_i^\dag = s_i^\dag \mathrm{e}^{i\pi \sum_{j<i}n_j}.
\end{equation}
We can decompose the bosonic operator into a density and a phase \cite{luther1974single, fisher1989boson}:
\begin{equation}\label{boson_of_phase}
    s_i^\dag = \sqrt{\rho_i} \mathrm{e}^{i\tilde{\theta}_i}.
\end{equation}
If we only consider low energy excitations, we can linearize the spectrum around the Fermi momenta and define left- and right-moving fermions according to the side of the spectrum at which they arise \cite{giamarchi2003quantum}. This corresponds to the description of the free fermion model as a Luttinger liquid \cite{giamarchi2003quantum, haldane1981luttinger}. Passing to the continuum limit and using the relations \eqref{fermion_of_boson} and \eqref{boson_of_phase}, while changing the sum to an integral over an infinite chain, we can write for the left- and right-moving fermions upon linearizing the spectrum \cite{fisher1989boson, miranda2003introduction}:
\begin{equation}\label{continuous_operators}
    c^\dag_{R/L}(x) = \frac{c_{j,R/L}^\dag }{\sqrt{a}} \approx \frac{1}{\sqrt{a}}\mathrm{e}^{i\tilde{\theta}(x)}\mathrm{e}^{\pm i\pi \int_{-\infty}^x \rho(x) dx}.
\end{equation}
We decompose the continuous density operator into a mean and fluctuations $\rho(x) = (\rho_0 + \tilde{\rho})$ with $\rho_0 = a k_F/\pi$. Quantum mechanics imposes the commutation relation $[\tilde{\theta}(x),\tilde{\rho}(y)]=i\delta(x-y)$ between the density and the phase \cite{fisher1989boson}. We can write
\begin{equation}\label{density_normalization}
    \rho(x) = \rho_0 + \frac{\partial_x \tilde{\phi}(x)}{\pi},
\end{equation}
where we introduced the field $\phi(x)$ by $\tilde{\rho} = \partial_x \tilde{\phi}(x)/\pi$ (we introduce the phase $\tilde{\phi}$ and accordingly $\tilde{\theta}$ to distinguish with the phase $\phi$ from Eqs. \eqref{first_currents}). Then the above commutation relation is achieved for
$[\tilde{\theta}(x),\tilde{\phi}(y)] = i\frac{\pi}{2} \text{sgn} (x-y)$. Plugging the form of $\tilde{\rho}$ into Eqs. \eqref{continuous_operators} gives upon accounting for the normalization imposed by Eq. \eqref{density_normalization}
\begin{subequations}\label{second_continuous_operators}
\begin{eqnarray}
    c^\dag_R(x) &=& \frac{1}{\sqrt{2\pi a}}\mathrm{e}^{i\tilde{\theta}(x)}\mathrm{e}^{i k_F x}\mathrm{e}^{i\tilde{\phi}}, \\
    c^\dag_L(x) &=& \frac{1}{\sqrt{2 \pi a}}\mathrm{e}^{i\tilde{\theta}(x)}\mathrm{e}^{-i k_F x}\mathrm{e}^{-i\tilde{\phi}}.
\end{eqnarray}
\end{subequations}
In order to retain fermionic commutation relations, we need to multiply the right-hand sides of the equations \eqref{second_continuous_operators} by the respective Klein factors $U_{R/L}$ where $U_R U_L = i$ \cite{miranda2003introduction}.
The total density is
\begin{equation}
    \rho(x) = \rho_0 + \frac{\partial_x \tilde{\phi}}{\pi} = c_R^\dag c_R + c_L^\dag c_L + \mathrm{e}^{i2k_F x} c_R^\dag c_L +\mathrm{e}^{-i2k_F x} c_L^\dag c_R,
\end{equation}
and we can write the density fluctuations as
\begin{equation}
    \frac{\partial_x \tilde{\phi}}{\pi} = \lim_{a\rightarrow 0} c_R^\dag(x+a) c_R(x) + c_L^\dag(x-a) c_L (x).
\end{equation}

Regarding the hopping part of the initial Hamiltonian \eqref{interacting_fermion_hamilton} and going to the continuum limit, we get
\begin{equation}
    H = -2{J}_{xy} \int dx c^\dag (x) c(x+a) +\text{h.c}.
\end{equation}
Splitting into the right- and left-moving branches and taking the limit of $a \rightarrow 0 $, we obtain after an integration by parts the Hamiltonian in the Dirac-form,
\begin{equation}
    H = i v_0 \int dx (c_R^\dag(x) \nabla c_R(x) - c_L^\dag(x) \nabla c_L(x)),
\end{equation}
with $v_0 = 4 a J_{xy}$.
The Hamiltonian in terms of the fields $\tilde{\phi}$ and $\tilde{\theta}$ reads
\begin{equation}
    H = \frac{v_0}{2\pi} \int dx ((\partial_x \tilde{\theta})^2 + (\partial_x \tilde{\phi})^2).
\end{equation}
The interaction term in Eq. \eqref{interacting_fermion_hamilton} takes the form $4 J_z \sum_i \left(n_i - \frac{1}{2} \right) \left( n_{i+1} - \frac{1}{2} \right)$. As we consider the system at half-filling, we can replace terms like $n_i - \frac{1}{2}$ directly by the density fluctuations and then write in the continuum limit and omitting the rapidly oscillating parts:
\begin{equation}
  a 4 J_z \int \left( \frac{\partial_x \tilde{\phi}}{\pi}\right)^2 = \frac{a 8 J_z / \pi}{2\pi} \int dx \left( \partial_x \tilde{\phi} \right)^2,
\end{equation}
so that the interacting Hamiltonian can be written upon introducing the Luttinger parameter $K$ and renormalizing the velocity $v$ as
\begin{equation}\label{interacting_Hamiltonian}
    H = \frac{v}{2\pi} \int dx \frac{1}{K} (\partial_x \tilde{\phi})^2 + K(\partial_x \tilde{\theta})^2,
\end{equation}
with $v/a = \sqrt{(4J_{xy})^2+32 J_z J_{xy}/\pi}$ and $K = \sqrt{4 J_{xy}/(4{J}_{xy}+8 J_z/\pi)}$.\par

Now, we include the disorder term $\frac{U_{af}}{2}\sum_j \sigma_j^z \tau_j^z$ with a staggered configuration $\tau_j^z = (-1)^j$. In the fermionic language this reads
\begin{equation}
  \frac{U_{af}}{2} \sum_j (-1)^j 2 n_j,
\end{equation}
where we neglected a constant.
In the continuum limit, we write $(-1)^j \rightarrow \mathrm{e}^{i\pi x/a}$ with $x = aj$ and thus obtain
\begin{eqnarray}
    \frac{U_{af}}{2a} \int dx \mathrm{e}^{i\pi \frac{x}{a}} (c_L^\dag c_L + c_R^\dag c_R) \nonumber \\+ \frac{U_{af}}{2\pi a} \int dx \, \mathrm{e}^{i\pi \frac{x}{a}}  i\mathrm{e}^{i2k_F x} \mathrm{e}^{i2\tilde{\phi}} + \text{h.c}.
\end{eqnarray}
The first integral can be neglected as it is oscillating rapidly. Since $k_F = \frac{\pi}{2a}$ and $x=aj$,
\begin{equation}
  \mathrm{e}^{i\pi \frac{x}{a}} \mathrm{e}^{i2k_F x} = (-1)^2 = 1
\end{equation}
and we can write the second integral as
\begin{equation}
  -\frac{U_{af}}{ \pi a}  \int dx  \sin(2\tilde{\phi}).
\end{equation}
Then the full Hamiltonian reads
\begin{widetext}
\begin{equation}\label{sine_gordon_Hamiltonian}
    H = \frac{v}{2\pi} \int dx \left( \frac{1}{K} (\partial_x \tilde{\phi})^2+K(\partial_x \tilde{\theta})^2\right) -  \frac{1}{\pi}\int dx \frac{U_{af}}{a} \sin(2\tilde{\phi}).
\end{equation}
\end{widetext}
From Eq. \eqref{sine_gordon_Hamiltonian}, it is evident that if $U_{af}$ is large, the sine-Gordon term dominates and we can anticipate a pinning of $\phi$ to $\pi/2$, so the system acquires a gap in the energy spectrum. Hereafter, we will look at the opposite case of small $U_{af}$, such that we can treat the sine-Gordon term as a perturbation. The pinning of the phase $\tilde{\phi}$ also engenders an exponential suppression of the current through the conjugate phase $\tilde{\theta}$.

\subsection{Renormalization group analysis}\label{RG_analysis}
\label{RG}

From now on, we assume that $U_{af} \ll J_{xy}$ so that we can do a perturbative analysis in the matter-impurities interaction. Our objective here is to write down the Renormalization Group (RG) equation for $U_{af}$ using the standard methodology \cite{giamarchi2003quantum}.
Assume we change the lattice parameter $a \rightarrow a' = a \mathrm{e}^{dl} \approx a(1+dl)$. From this it follows that $dl = \log (a'/a)$. We demand that the partition function remains unchanged under this transformation, i.e. $Z(a') = Z(a)$.
Here, this gives the equation:
\begin{equation}
    \frac{U_{af}^2(a)}{a^2} a^{2K} =  \frac{U_{af}^2(a')}{a'^2} a'^{2K}.
\end{equation}
It is useful to redefine the dimensionless quantity $g_{af} = U_{af} a/v$ such that
\begin{equation*}
    g_{af}^2 a^{2K-4} =  g_{af}^2(a')a'^{2K-4}.
\end{equation*}
Upon scaling the lattice constant $a' = a \mathrm{e}^{dl}$ we obtain
\begin{align}\label{RG_equation_Uaf}
    \dv{g_{af}}{l} &= (2-K)g_{af}.
\end{align}
Now for the simple case of $J_z = 0$ we have $K = 1$. Therefore, upon increasing $l$, we also enhance $g_{af}$. Define $g_{af}(l^*)$, at which this term is for general $K$ of the same order as the hopping term.
Solving the differential equation \eqref{RG_equation_Uaf} by integrating from $a$ to $l$ we get
\begin{equation}
    \frac{g_{af}(l)}{g_{af}(a)} = \left( \frac{a(l)}{a} \right)^{2-K}.
\end{equation}
We fix $l^*$ at which the impurities-matter term is strongly renormalized and becomes comparable to the kinetic energy, which gives
\begin{equation}
    g_{af}(l^*) = \frac{J_{xy}a}{v},
\end{equation}
and therefore:
\begin{equation}\label{l_star}
    l^* \sim a\left( \frac{v}{a U_{af}} \right)^{\frac{1}{2-K}}.
\end{equation}
\par
In fact, for weak Gaussian disorder, an RG equation similar to \eqref{RG_equation_Uaf} can be found following Ref. \cite{giamarchi2003quantum} and reads translated to our setup
\begin{align}
    \dv{g_{af}}{l} &= (3-2K)g_{af},
\end{align}
leading to an estimation of the localization length
\begin{equation}\label{l_star_gia}
    l^* \sim  a\left( \frac{v}{a U_{af}} \right)^{\frac{1}{3-2K}}.
\end{equation}
The expressions \eqref{l_star} and \eqref{l_star_gia} quantify the gap opened by a sine-Gordon term, which arises from a staggered magnetic field or a Gaussian disorder in the original model. It is interesting to observe that for $J_z=0$ the localization lengths for the setup with the telegraph potential and with Gaussian disorder are very similar which supports our conclusion that a well-pronounced localization occurs for various forms of random potentials in the strongly-coupled rungs regime.

\begin{figure}
  \centering
  \includegraphics[width=\linewidth]{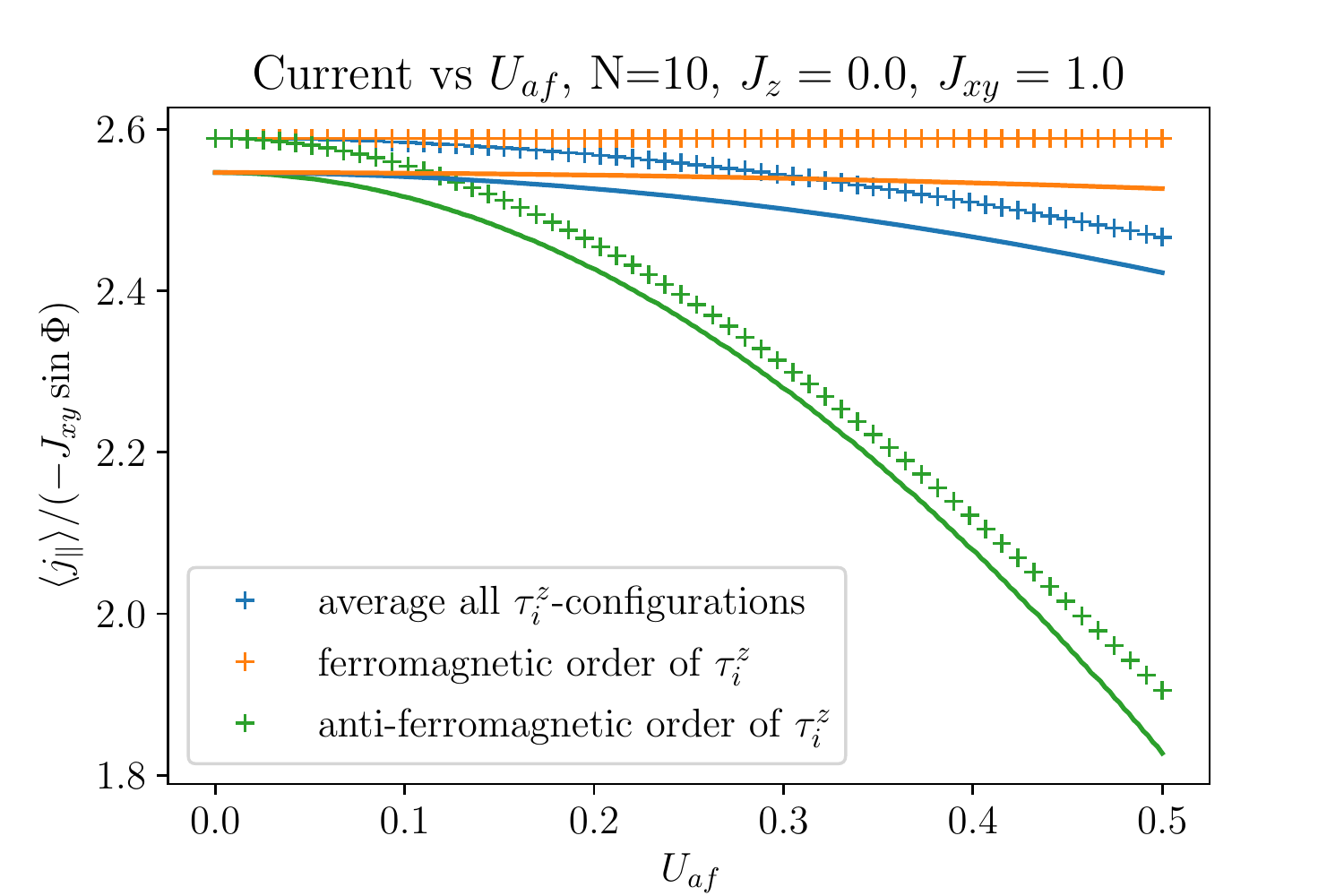}
  \captionof{figure}{A comparison of results from ED for ferromagnetic, antiferromagnetic and average over all $\tau_i^z$-configurations with $\phi = 0.001$, $J_{xy}=1.0$, $J_z=0.0$ a system with $N=10$ sites and periodic boundary conditions.} The theoretical description \eqref{simple_current_theory} for a ferromagnetic configuration and the fitting of a truncated sum \eqref{current_sum_2} for the antiferromagnetic and averaged setup respectively are shown as solid lines. To obtain the fit, we sum over 4000 states to obtain a smooth curve. The fitting parameter $C$ in Eq. \eqref{fit_equation} was evaluated to $C = 0.1624$ for the antiferromagnetic configuration and to $C=0.3993$ for the average over all disorder configurations.
  Details on the numerical implementation and the fitting procedure can be found in Appendix \ref{numerics}.
	\label{localization_figure}
\end{figure}

For $J_z=0$ and without disorder, the spectrum can be found in momentum space in a similar fashion as Eq. \eqref{Hamiltonian_diagonal}
\begin{equation}\label{Hamiltonian_diagonal_free}
    H = -4{J}_{xy}\sum_k \cos(a k-\phi) c_k^\dag c_k,
\end{equation}
from which the current is easily evaluated as
\begin{equation}\label{current_sum_2}
    j = \frac{2}{N}\pdv{H}{\phi} = -4{J}_{xy} \frac{2}{N} \sum_k \sin{(a k-\phi)}c_k^\dag c_k.
\end{equation}
In both equations, the sum ranges over accessible momentum states, spaced depending on the boundary conditions (see the discussion in Sec. \ref{ferro_order_section}) up to the Fermi momenta $k_F = \pi/(2a)$.
Disorder will open a gap quantified by $l^*$. We therefore account for its effect by introducing a cut-off of the sum \eqref{current_sum_2} so that it ranges only over momenta with absolute values smaller than $| k_F - (l^*)^{-1}|$. This can be done numerically to obtain a prediction for the behaviour of the current with a staggered magnetic field and the disorder averaged current expectation value. The approximative nature of Eqs. \eqref{l_star} and \eqref{l_star_gia} can be accounted for by fitting a free prefactor $C$, i.e. for $K=1$ we fit $C$ in
\begin{equation}\label{fit_equation}
l^* = C a\left( \frac{v}{a U_{af}} \right).
\end{equation}
In this case, up to this prefactor both cases show the same effects for small $U_{af}$. This can be confirmed for a small system using numerics, which is shown in Fig. \ref{localization_figure}. Here, the orange and the green crosses show the current expectation value for a ferromagnetic and an anti-ferromagnetic configuration of the $\tau_i^z$-variables respectively. The blue crosses show an average of the current expectation value over all possible configurations of $\tau_i^z$. Since in Fig. \ref{localization_figure} $J_z = 0$, the ED results are the same as in Fig. \ref{ED_comparison_ordered}, but with a smaller range of $U_{af}$.
We see in Fig. \ref{localization_figure} that for $U_{af} = 0$ there is an offset between the ED results and the theoretical curves. For the orange curve representing Eq. \eqref{simple_current_theory} and the orange crosses this can be ascribed to the fact that the ED data was obtained for a small system ($N=10$), while in the derivation of Eq. \eqref{simple_current_theory} we replaced the summation by an integration between the two occupied states closest to the Fermi momentum (see Fig. \ref{band}).
In a similar way, for the truncated summation over momentum states shown by the blue and green curves representing equation \eqref{current_sum_2}, we took a large number of sites in order to obtain a smooth curve. Additionally, for the bosonization approach described in this Section, we linearized the spectrum in order to obtain Eq. \eqref{continuous_operators}. \par
We see that Eq. \eqref{alternating_current_sum} could account for the full range of $U_{af}$ values in agreement with ED results, while the approach described here holds only for small values of $U_{af}$. The advantage of the latter is however, that thanks to the bosonized framework in which we developed it, it is applicable also in the presence of Ising interactions proportional to $J_z$. We will exploit this in the following.\par
\begin{figure}
	\centering
	\includegraphics[width=\linewidth]{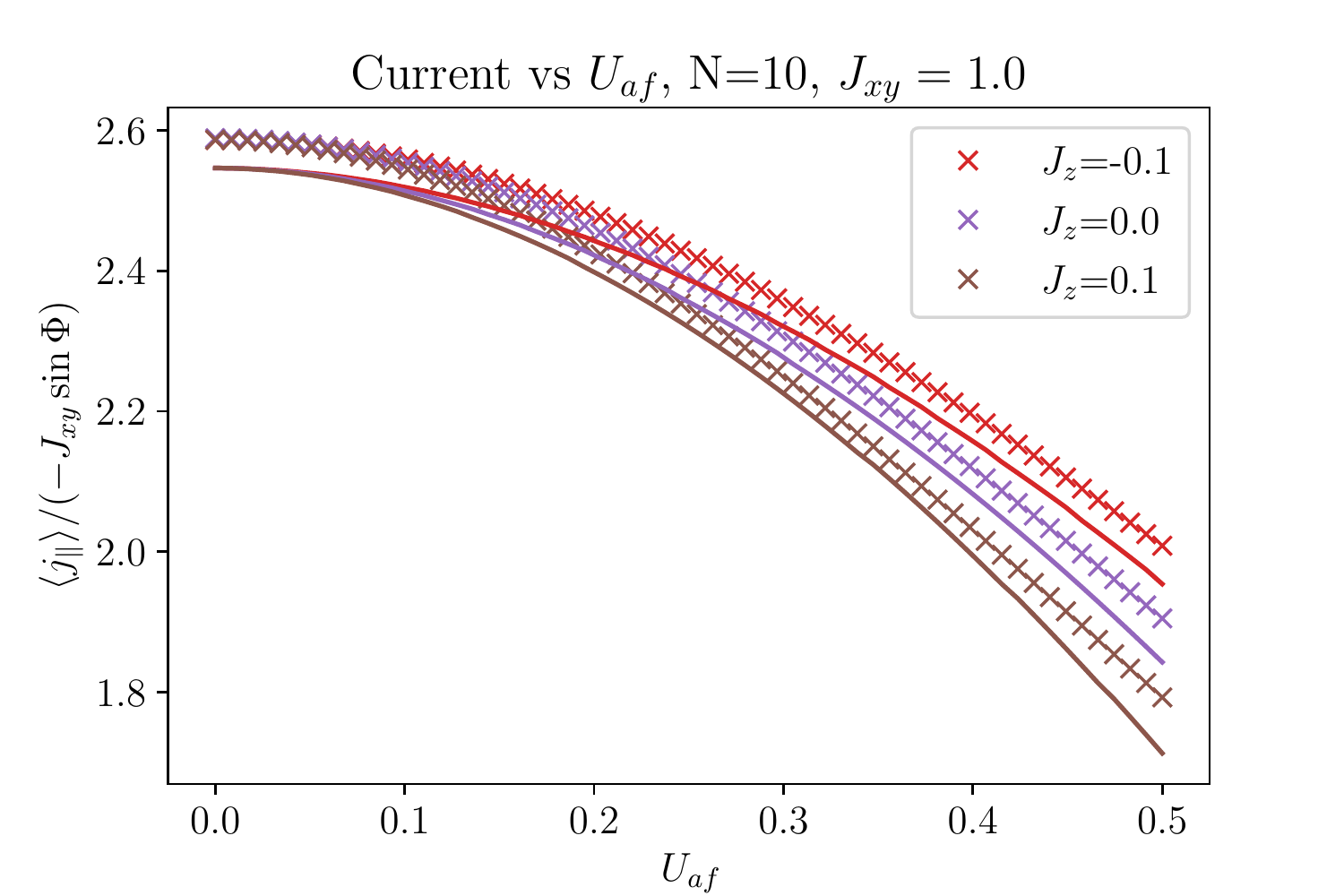}
  \captionof{figure}{Current from ED as a function of $U_{af}$ with a staggered magnetic field (anti-ferromagnetic configuration of $\tau_i^z$) in comparison to the sum \eqref{current_sum_2} truncated at $| k_F - (l^*)^{-1}|$ (fitted for $C$). The results are shown for the non-interacting case and for a small positive and negative interaction $J_z = \pm 0.1$. We set $J_{xy} = 1.0$, $g = 0.0$ and $\phi = 0.001$ with periodic boundary conditions. The ED data (crosses) was obtained for a chain with ten sites. The truncated sum (solid lines) was evaluated on a large system with scaled lattice constant in order to get a smooth result (see Appendix \ref{numerics} for details) and using a fitting for $C$.}
	\label{staggered_localization}
\end{figure}

As we see from Eq. \eqref{l_star}, decreasing $K$ (corresponding to an antiferromagnetic coupling $J_z>0$) leads to a decrease in $l^*$, therefore the gap increases and there are more terms which are cut off from the sum in Eq. $\eqref{current_sum_2}$. This effect is stronger than the increase in $v$ coming from an antiferromagnetic coupling $J_z$, as can be seen from simulation results in Fig. \ref{staggered_localization}.
This means that an antiferromagnetic coupling of the $\sigma_i^z$-spins in this regime supports the localization of the current, which is seen in Fig. \ref{staggered_localization} from ED and from the bosonization approach for an antiferromagnetic configuration of the disorder. A ferromagnetic coupling ($J_z<0$) tends to hinder the localization in this setup, which is in that sense also consistent with the result from the bosonization approach.

\par

\begin{figure}
	\centering
	\includegraphics[width=\linewidth]{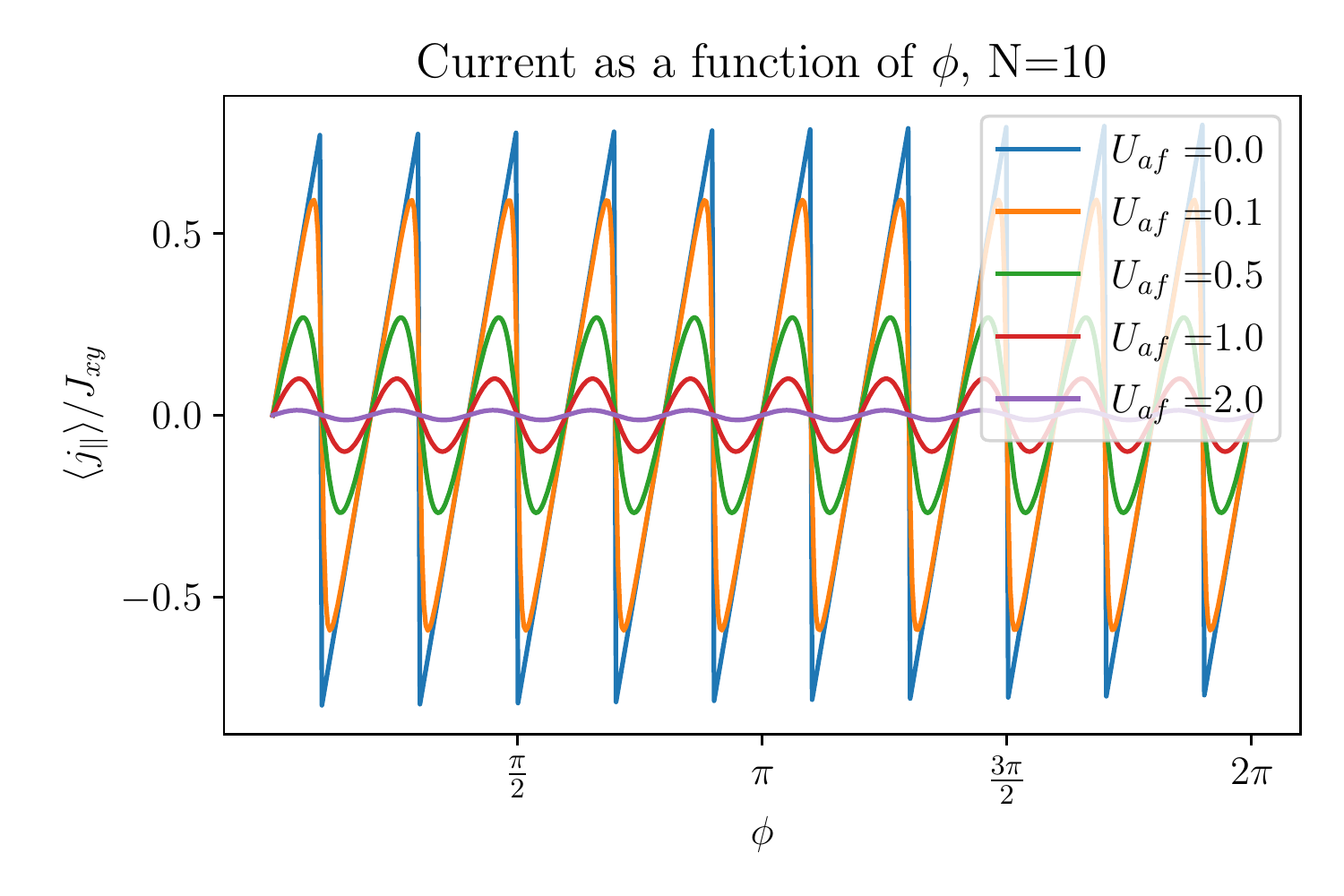}
  \captionof{figure}{Current for an antiferromagnetic configuration of $\tau_i^z$ in a system with $N=10$ sites, result from ED as a function of $\phi$ and for a range of $U_{af}$ values.
  We set $J_{xy} = 1.0$, $J_z=0.0$ and $g = 0.0$ with periodic boundary conditions.}
	\label{current_phi_dependence}
\end{figure}\par
To end this section, we remark that for our evaluation of the localization length $l^*$ we assumed for simplicity that $\phi \rightarrow 0$. To include the effects of a finite $\phi$ perturbatively, one has to add back the Peierls phases to the hopping part of Hamiltonian \eqref{interacting_fermion_hamilton} and rewrite them in terms of the sine and the cosine of $\phi$.
Considering the cosine part only, we see that this changes $J_{xy} \rightarrow J_{xy} \cos\phi$, so for small phases one can consider the correction $J_{xy} \rightarrow J_{xy} (1-\phi^2)$. This would modify both $v$ and $K$.
On a more general note, for the small systems considered and with an antiferromagnetic configuration of the $\tau_i^z$-variables the current depends on $\phi$ in a periodic way depending on the number of sites. This is shown for $J_z = 0$ in Fig. \ref{current_phi_dependence}. The ED result presented there agrees with Eq. \eqref{alternating_current_sum}, from which also the dependence of period on the number of available momentum states and thereby on the number of sites can be understood.
\par
In the following, we will discuss the possibility of a many-body localized phase for strong disorder.

\section{Many-Body Localization}
\label{MBL}

We emphasize here the recent interest at the interface between localization effects and gauge theories and especially in quantum spin models \cite{singh2016signatures,pal2010many,FreeLoc,FreeLoc2,Moore}. We will then study the link with many-body localization starting directly
from the XXZ model in Eq. (\ref{effective_XXZ}) for the specific situation of a two-peak random potential. It is also relevant to mention here that many-body physics related to two-fluids models has recently been studied in the specific situations of two-peak or binary random potentials with possible applications in cold atoms \cite{andraschko2014purification, tang2015quantum}. From Refs. \cite{singh2016signatures} and \cite{pal2010many}, a disorderded potential drawn from a box distribution and $J_z = -J_{xy}$ drives a many-body localized phase for $g=0$. In our case, we are considering a peaked disorder and stability of this phase upon adding a small transverse field $g$. To make a link with the results in Ref. \cite{singh2016signatures}, we will study the entanglement entropy and bipartite fluctuations upon tracing a region of the system \cite{song2012bipartite}. In Ref. \cite{vznidarivc2008many} the entanglement entropy has also been used as an indicator for localization in an XXZ-model for different values of the interaction $J_z$ with a disorderded potential drawn from a box distribution and with antiferromagnetically ordered impurities.\par
We will first present some results for the bipartite fluctuation in the ground state in the decoupled rung limit with large $g$ (Sec. \ref{weakly_coupled_fluctuation}) and with $g=0$ (Sec. \ref{XXZ_fluctuation}).
In order to detect a many-body localized phase, we will finally depart from the ground state and consider the time evolution after a quench from an initially prepared pure state of the $\vec{\sigma}$-spins in the rung-Mott phase, which in our case will be the N\'eel state in $z$-direction (Sec. \ref{longTimeEvolution}). There we will consider $g=0$ and the effect of a small value for $g$.
\par
The bipartite fluctuations of the spin in a state $\ket{\psi}$ read for a subsystem of size $l$ \cite{song2012bipartite}
\begin{equation}\label{bipartite_fluctuation}
  \mathcal{F}(l) = \bra{\psi} (S_{l}^z)^2 \ket{\psi} - \bra{\psi} S_{l}^z \ket{\psi} ^2,
\end{equation}
with $S_{l}^z = \sum_{i=1}^{l} \frac{1}{2}\sigma_i^z$. First, for a comparison we address the situation of the weakly-coupled rungs limit where we can also solve the dynamics.

\subsection{Weakly-Coupled Rungs limit}\label{weakly_coupled_fluctuation}

In the weakly-coupled rungs limit with $g \gg J_{xy}, J_z$, in the ground state evaluated as in \ref{static_impurities} we can readily calculate the bipartite fluctuation in the ground state which reads
\begin{equation}\label{weakly_coupled_F}
    \mathcal{F}(l) = \frac{l}{4} \frac{1}{1+(U_{af}/2g)^2}.
\end{equation}

\begin{figure}[b]
  \includegraphics[width = \linewidth]{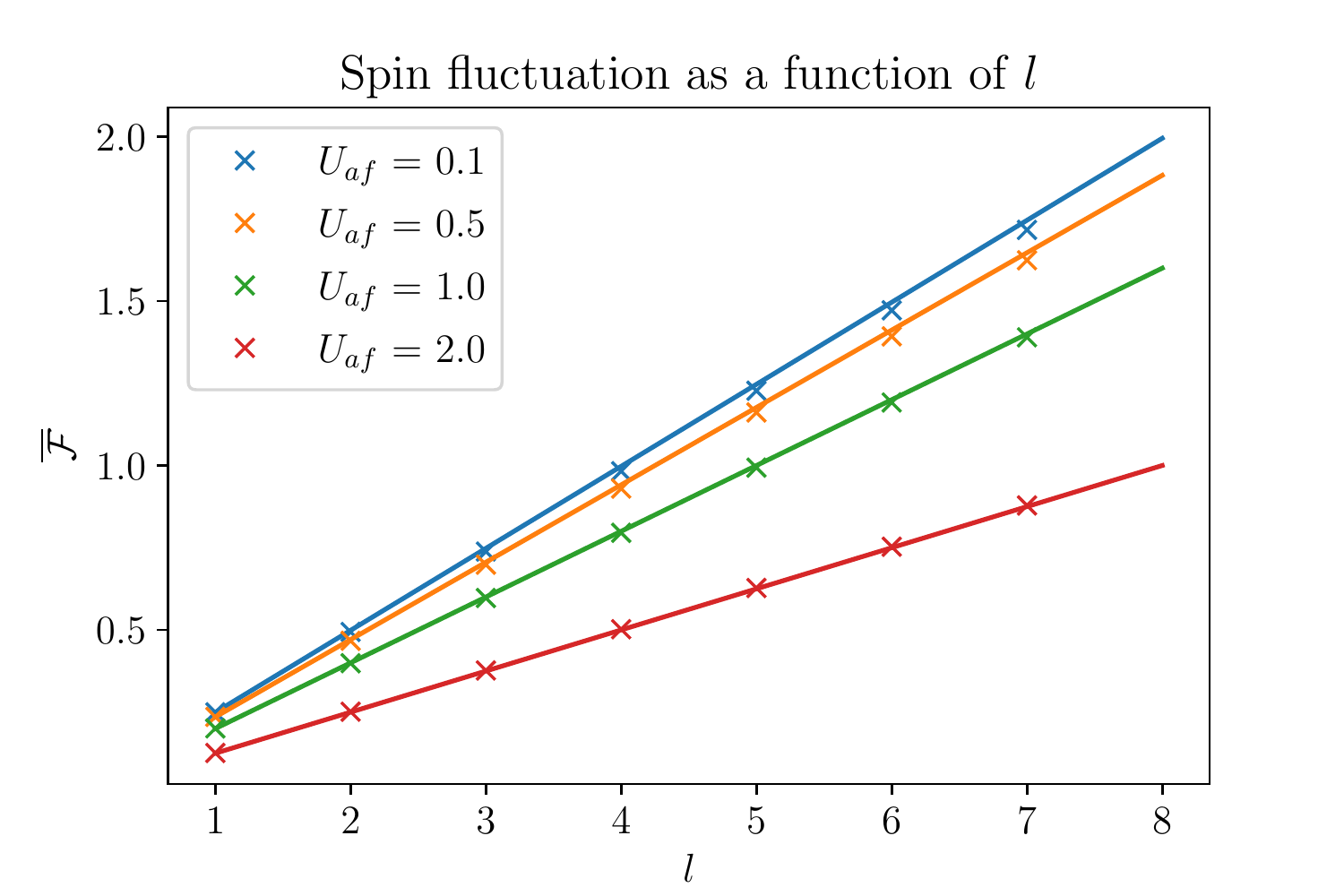}
  \caption{ED results (crosses) for the bipartite fluctuation as a function of the subsystem size in the weakly-coupled rungs limit for a range of $U_{af}$ values. The size of the full system is fixed to $N=8$ and we use $J_{xy}=J_z=0.01$, $g=1.0$, $\phi = 0.01$ and open boundary conditions. The bipartite fluctuation $\overline{\mathcal{F}}$ was evaluated as an average over all possible configurations of $\tau_i^z$. The solid line shows a comparison to Eq. \eqref{weakly_coupled_F}. The bipartite fluctuation grows linearly with the subsystem size.}
 	\label{bipartite_spin_decoupled_L}
\end{figure}

Up to a prefactor, it is equal to the parallel current and it scales linearly with the subsystem size.
This is independent of the disorder configuration since the localization occurs on each rung independently. In Fig. \ref{bipartite_spin_decoupled_L}, we show the scaling of the bipartite fluctuation with the subsystem size as an average over disorder configurations which we call $\bar{\mathcal{F}}$. In this Section, expectation values are implicit through the definition \eqref{bipartite_fluctuation} of the bipartite fluctuation.

\subsection{XXZ-chain}\label{XXZ_fluctuation}

\begin{figure}[b]
  \includegraphics[width = \linewidth]{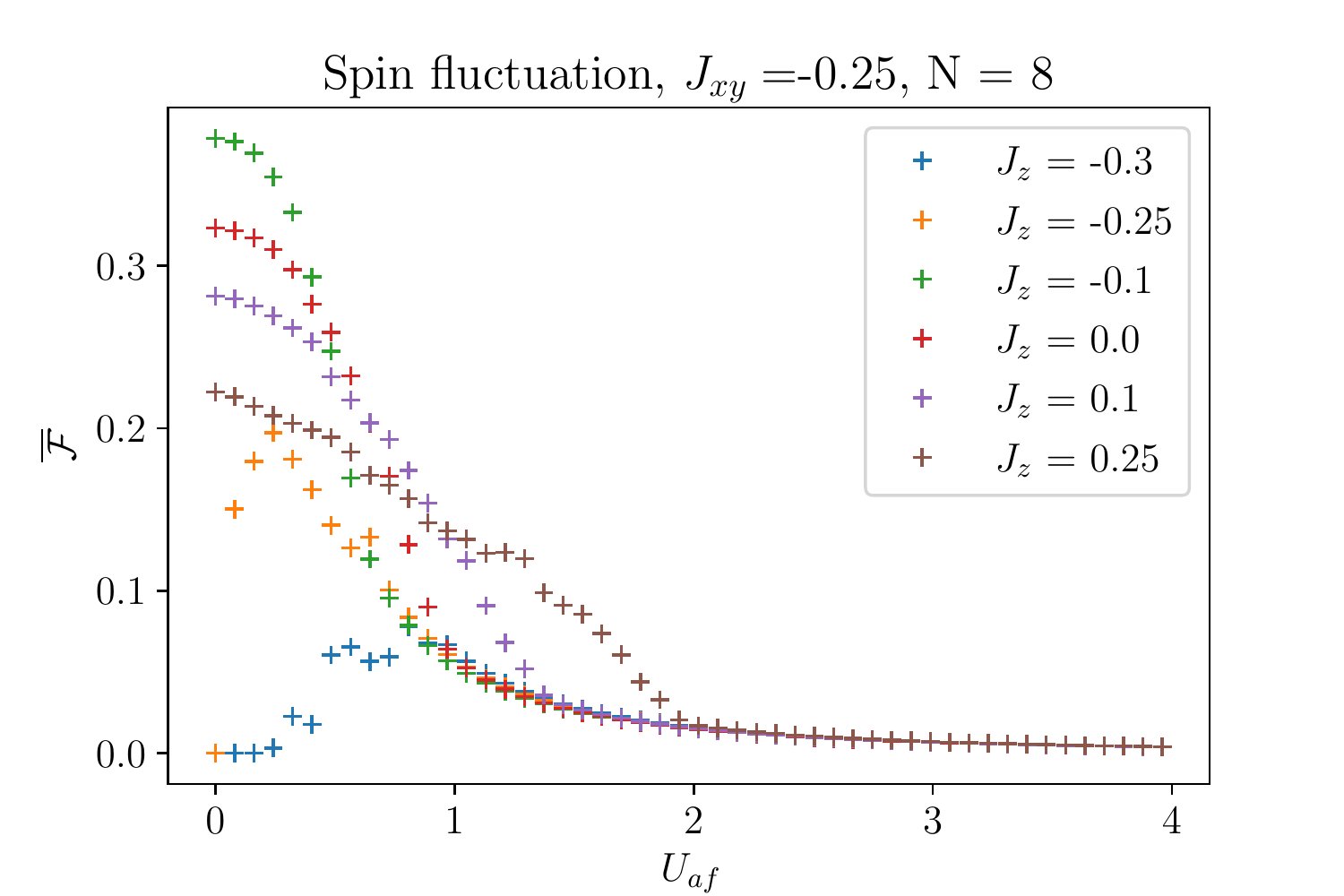}
  \caption{Bipartite fluctuations averaged over all realizations of $\tau_i^z$ for $N=8$ with periodic boundary conditions, the bipartition boundary being in the center of the chain (i.e. $l = N/2$). Here, we set $J_{xy} = -0.25$ and $g=0.0$, so that the curve with $J_z = 0.25$ bridges with the results in \cite{singh2016signatures}. We furthermore set $\phi = 0.001$.}
 	\label{bipartite_spin_N8}
\end{figure}

When deriving the Hamiltonian \eqref{effective_spin} from the bosonic ladder model (see Sec. \ref{rungMottDefinitions}), we have $J_{xy}>0$ or $-J_{xy}<0$.
Here, we redefine $J_{xy}\rightarrow -J_{xy}$ in Eq. \eqref{effective_XXZ} corresponding now to ferromagnetic transverse spin $J_{xy}$ couplings such that we can apply results from Ref. \cite{song2010general} and compare in Sec. \ref{longTimeEvolution} with results from Ref. \cite{singh2016signatures}. Note that both cases can be mapped onto each other by rotating around the z-axis for every second spin.
In the model of Eq. \eqref{effective_spin}, we can set $J_{xy} = -0.25$, $J_z = 0.25$ and $g=0$ to make the connection with Ref. \cite{singh2016signatures}. Here, we study the bipartite fluctuations and entanglement measures when tracing half of the system, from the ground state, which will enables us to compare with the situation of a quench studied in the next section. In the ground state, the bipartite fluctuations can then be evaluated numerically and for a range of $J_z$ we obtain the results shown in Fig. \ref{bipartite_spin_N8}.

\begin{figure}[h]
  \includegraphics[width = \linewidth]{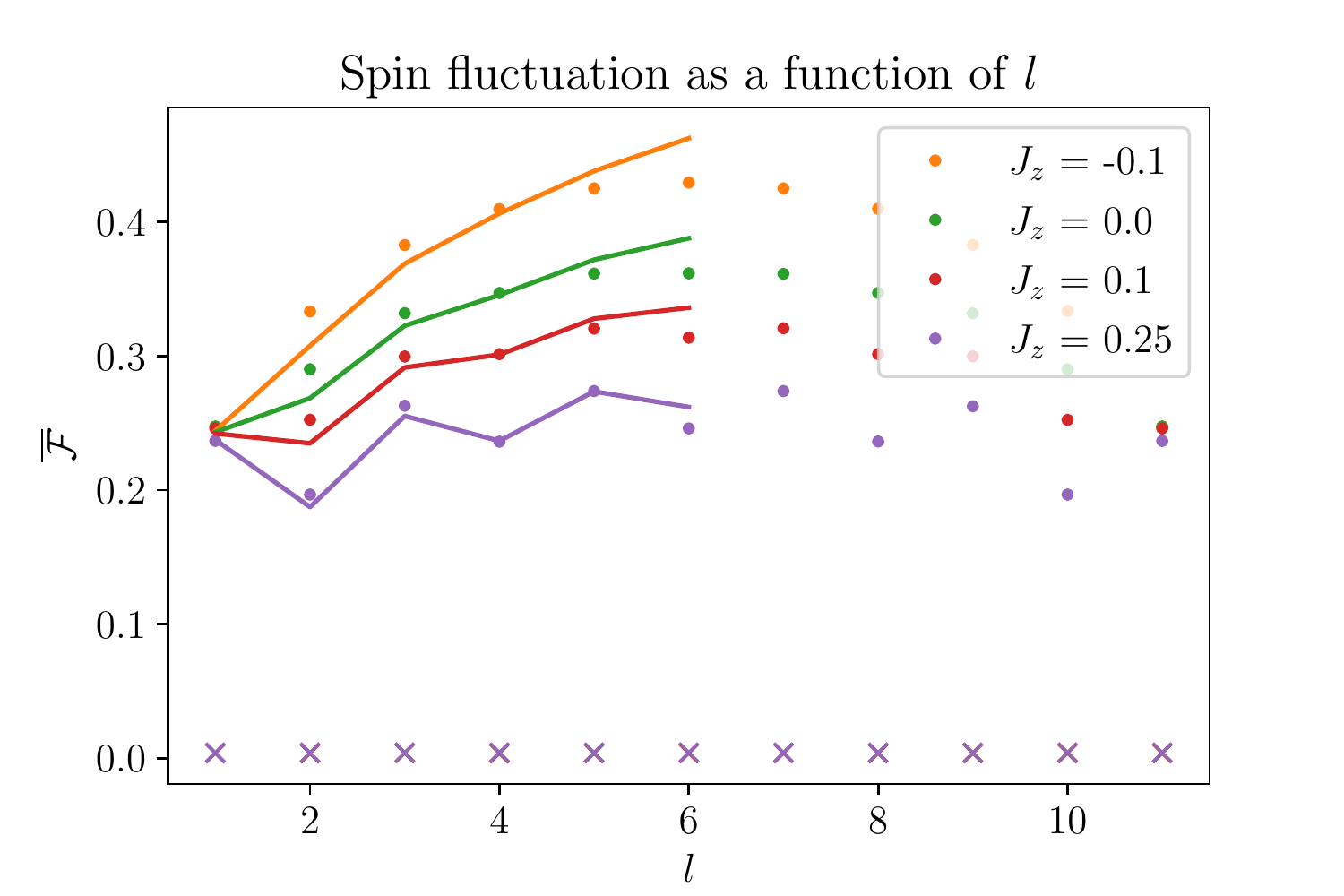}
  \caption{Scaling of the average bipartite fluctuations with the subsytem size for different values of $J_z$ and $U_{af}$. The dots show simulation results in the gapless phase with $U_{af} = 0.1$. The crosses show results for $U_{af} = 4.0$. Different values of $J_z$ are represented by different colors. The bipartite fluctuation was evaluated for 1000 randomly chosen configurations of $\tau_i^z$ and consequently averaged over. These simulations were performed for a chain with $12$ sites and periodic boundary conditions. We set $J_{xy} = -0.25$, $g=0.0$ and $\phi = 0.001$. For the results in the gapless phase (dots), we fit equation \eqref{bipartite_scaling_XXZ}, the results are represented by solid lines.}
 	\label{fit_spin_fluctuation_vs_L}
\end{figure}
The gapless phase of the XXZ-model in Eq. \eqref{effective_XXZ} is realized for $-1 < - J_z/J_{xy}  \leq 1$
In this phase, the low energy physics without disorder is described by the Luttinger liquid theory \cite{giamarchi2003quantum}. In the region $-1 < - J_z/J_{xy}  < 1$, the scaling of the fluctuations with the subsystem size can be evaluated when $U_{af}\rightarrow 0$ and reads \cite{song2010general, song2012bipartite}
\begin{equation}\label{bipartite_scaling_XXZ}
  \mathcal{F}(l) = \frac{K}{\pi^2} \ln(l) + \frac{f_2}{\pi^2} - A_1 \frac{(-1)^l}{\pi^2 l^{2K}},
\end{equation}
where the Luttinger parameter $K$ is now determined from the Bethe ansatz solution
\begin{equation}\label{luttinger_parameter}
  K = \frac{1}{2}\left(1 - \frac{\cos^{-1}\Delta}{\pi} \right)^{-1},
\end{equation}
with here $\Delta = - J_z/J_{xy}$. The form in Eq. (\ref{interacting_Hamiltonian}) is only valid in the perturbative region $|J_z|\ll |J_{xy}|$. The $\ln l$ behavior comes from gapless modes in the effective fermions theory achieved by the Jordan-Wigner transformation. We will therefore refer to this phase at weak-coupling with impurities, i.e. starting from $U_{af}=0$, as gapless phase.  The $A_1$ term describes Friedel oscillations of the particle densities from the boundary. As can be seen from Eq. \eqref{luttinger_parameter} and the blue curve in Fig. \ref{bipartite_spin_N8}, the Luttinger parameter diverges for $\Delta \rightarrow -1$ traducing an instability or a gap opening for the sound modes fluctuations in the ferromagnetic region $J_z < J_{xy}$ (with $J_{xy}<0$) when $U_{af}=0$ \cite{giamarchi2003quantum, song2012bipartite}.
In that limit, Eq. (\ref{bipartite_scaling_XXZ}) is not valid. Instead, the ferromagnetic Ising phase shows a classical order and the bipartite fluctuations should then be zero when $U_{af}=0$. When switching on $U_{af}$, it is interesting to observe that certain configurations such as the antiferromagnetic situation for impurities produce quantum fluctuations restoring gapless modes for the fluctuations which leads upon averaging to the behaviour shown in Fig. \ref{fit_spin_fluctuation_vs_L} for $J_z=-0.25$ and $J_z=-0.3$.  Analytically, the occurrence of a gapless phase
at small $U_{af}$ can be understood using a mapping similar to the one of Sec. IV in Ref. \cite{Dmitriev} for the specific case of an antiferromagnetic ordering of the impurities. The Heisenberg point $\Delta = 1$ is also special since in the bosonized framework there is a marginal operator at this point so that the spin correlations acquire a correction \cite{giamarchi2003quantum, song2010general} and for $\Delta > 1$ the system becomes gapped and Eq. \eqref{bipartite_scaling_XXZ} needs to be modified \cite{giamarchi2003quantum, song2012bipartite}. \par
Including the effects of disorder through $U_{af}$, a phase transition to a localized phase is anticipated while for small values of disorder the scaling \eqref{bipartite_scaling_XXZ} of the bipartite fluctuation with subsystem size (for the disorder-free case) should still hold (at least qualitatively) as long as we are in the gapless phase with visible (bipartite) fluctuations. We verify that disorder or equivalently $U_{af}$ induces a transition to a localized phase by plotting the disorder-averaged bipartite fluctuation $\overline{\mathcal{F}}$ against the disorder strength $U_{af}$ for different values of $J_z$ in Fig. \ref{bipartite_spin_N8}. For $-1 < - J_z/J_{xy}  < 1$, the behaviour is qualitatively the same with a sharp transition to the localized phase for a critical value of disorder strength $U_{af}$ depending on $J_z$. The curve with $J_z = -0.25$ is in the ferromagnetic phase, consequently its behaviour is different for $U_{af} =0$ and the bipartite fluctuation vanishes. It is interesting to observe in Fig. \ref{bipartite_spin_N8} that for increasing values of disorder the behaviour of this curve can still be compared to the other curves qualitatively.\par

In the gapless phase (i.e. at small disorder), interestingly already for small systems sizes, we can fit the parameters $f_2$ and $A_1$ in Eq. \eqref{bipartite_scaling_XXZ} to the results from the simulation, which is shown in Fig. \ref{fit_spin_fluctuation_vs_L}.
The simulation results for the gapless phase are here represented by dots, the results from fitting to Eq. \eqref{bipartite_scaling_XXZ} for $U_{af} = 0.1$ are shown by solid lines.
In the strongly localized regime, results for a large value of disorder $U_{af}$ lead to a vanishing bipartite fluctuation for all values of $J_z$ (represented by crosses in the figure). For the considered situation, we observe that the curve with $J_z = 0.25$ corresponding to the Heisenberg point could be also fitted with Eq. \eqref{bipartite_scaling_XXZ} in Fig. \ref{fit_spin_fluctuation_vs_L} even though corrections are present in the form of the $A_1$ term \cite{song2010general}.

The entanglement entropy $S$ between two subsystems is defined from the von Neumann entropy
\begin{equation}\label{entanglement_entropy}
  S = -\Tr \rho_A \ln \rho_A,
\end{equation}
where $\rho_A = \Tr_B \left( \ket{\psi} \bra{\psi}\right)$ is the density matrix of the ground state with the degrees of freedom of the composite subsystem traced out. We have verified that the entanglement entropy of the half chain (i.e. for $l=N/2$) shows a similar transition as the bipartite fluctuations in Fig. \ref{bipartite_spin_N8}.

\subsection{Long time evolution}\label{longTimeEvolution}

We now turn towards a time-dependent protocole.
If we prepare the system of $\vec{\sigma}$-spins in the N\'eel state in $z$-direction $\ket{\uparrow \downarrow \uparrow \downarrow ... \uparrow \downarrow}$ and evolve in time with the Hamiltonian \eqref{effective_spin} with $J_{xy} = -0.25$, $J_z = 0.25$ and $\phi = 0.01$, we can evaluate the bipartite fluctuations and the entanglement entropy of the half chain (i.e. $l = N/2$) as an average over all configurations of disorder
(details on the numerical implementation can be found in appendix \ref{numerics}).
We can then qualitatively compare to the results obtained in \cite{singh2016signatures} for the time evolution of the bipartite fluctuations and the entanglement entropy.\par
For $g=0$, we observe a similar behaviour as with a box disorder in Ref. \cite{singh2016signatures}: With weak disorder, both the entanglement entropy and the bipartite fluctuation of the half chain saturate to a finite value after a rapid growth. In the localized phase, both the bipartite fluctuations and the entanglement entropy of the half chain show a rapid growth at short times, after which the bipartite fluctuations saturate and the entanglement entropy shows a logarithmic growth with time at strong interactions, in agreement with a many-body localization.\par
\begin{figure*}
\centering
 \includegraphics[width = 0.45\linewidth]{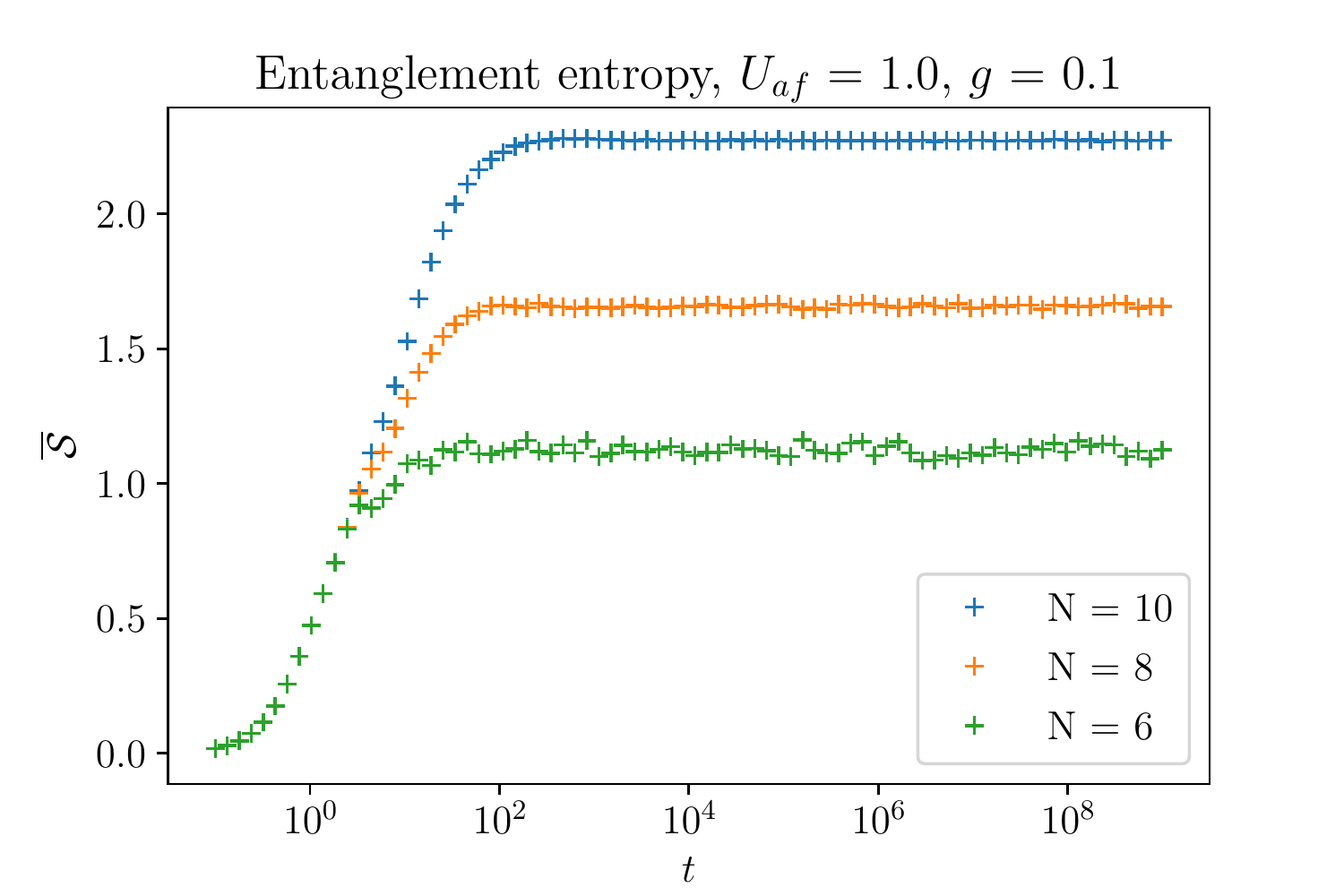}
 \includegraphics[width = 0.45\linewidth]{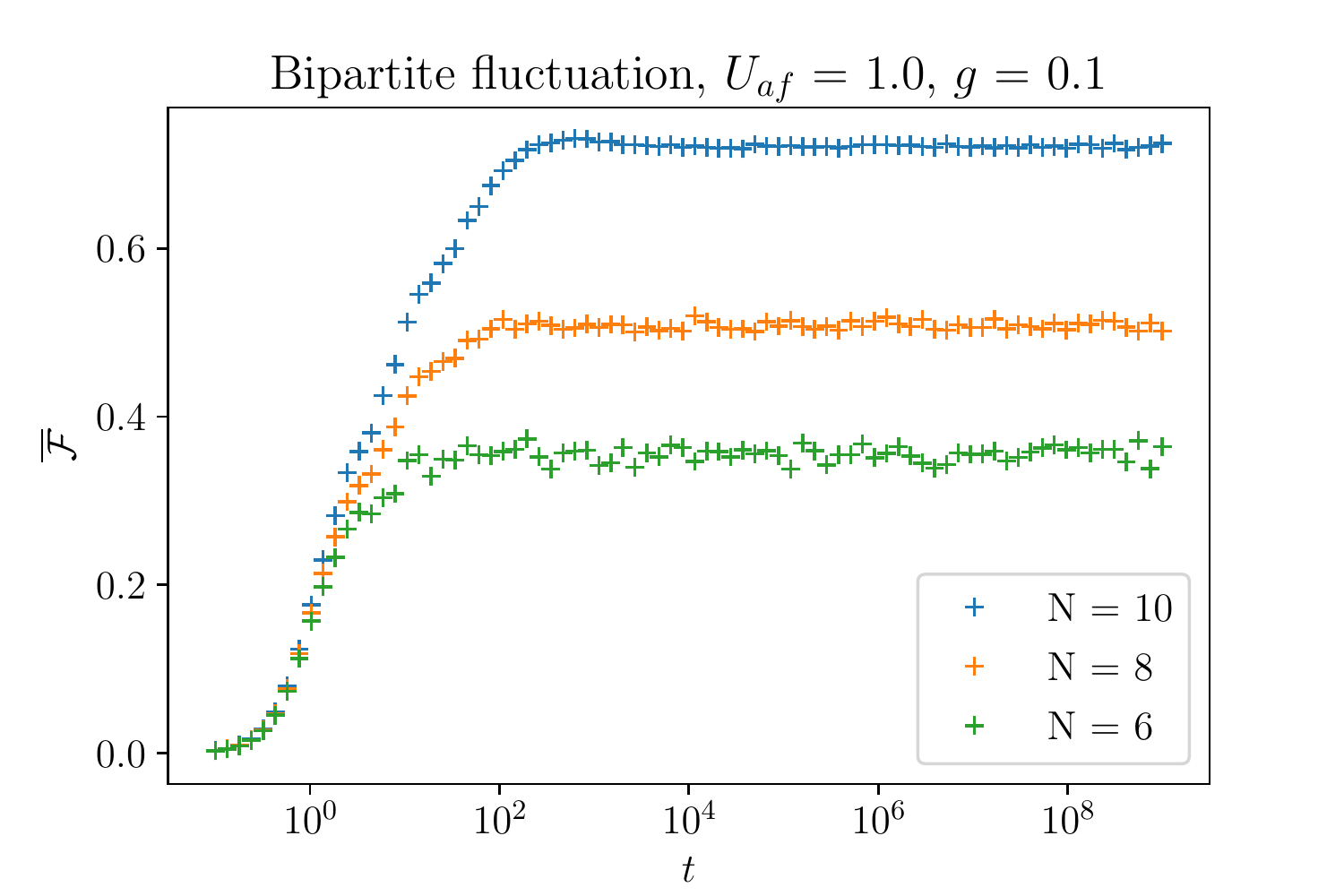}\\
 \includegraphics[width = 0.45\linewidth]{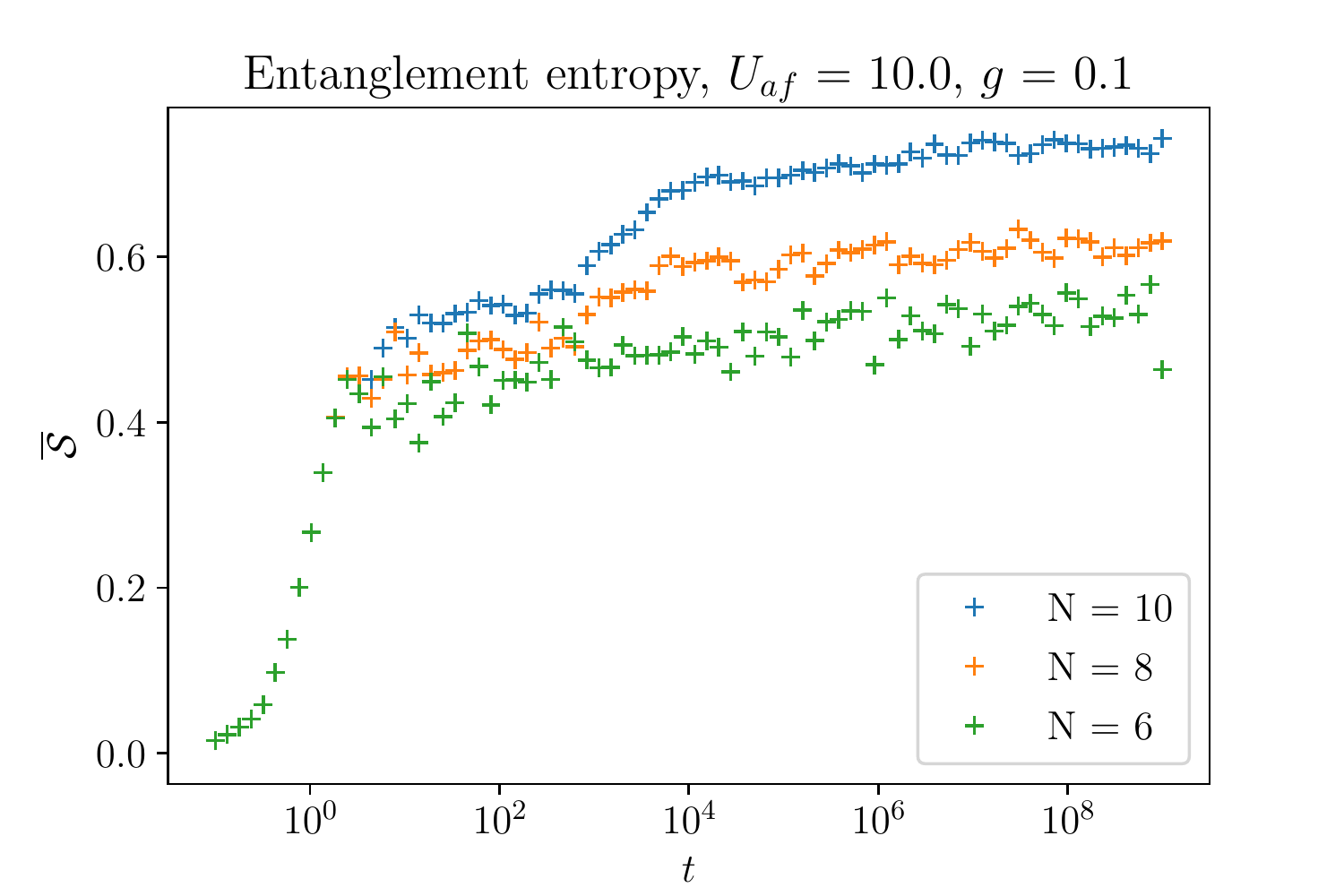}
 \includegraphics[width = 0.45\linewidth]{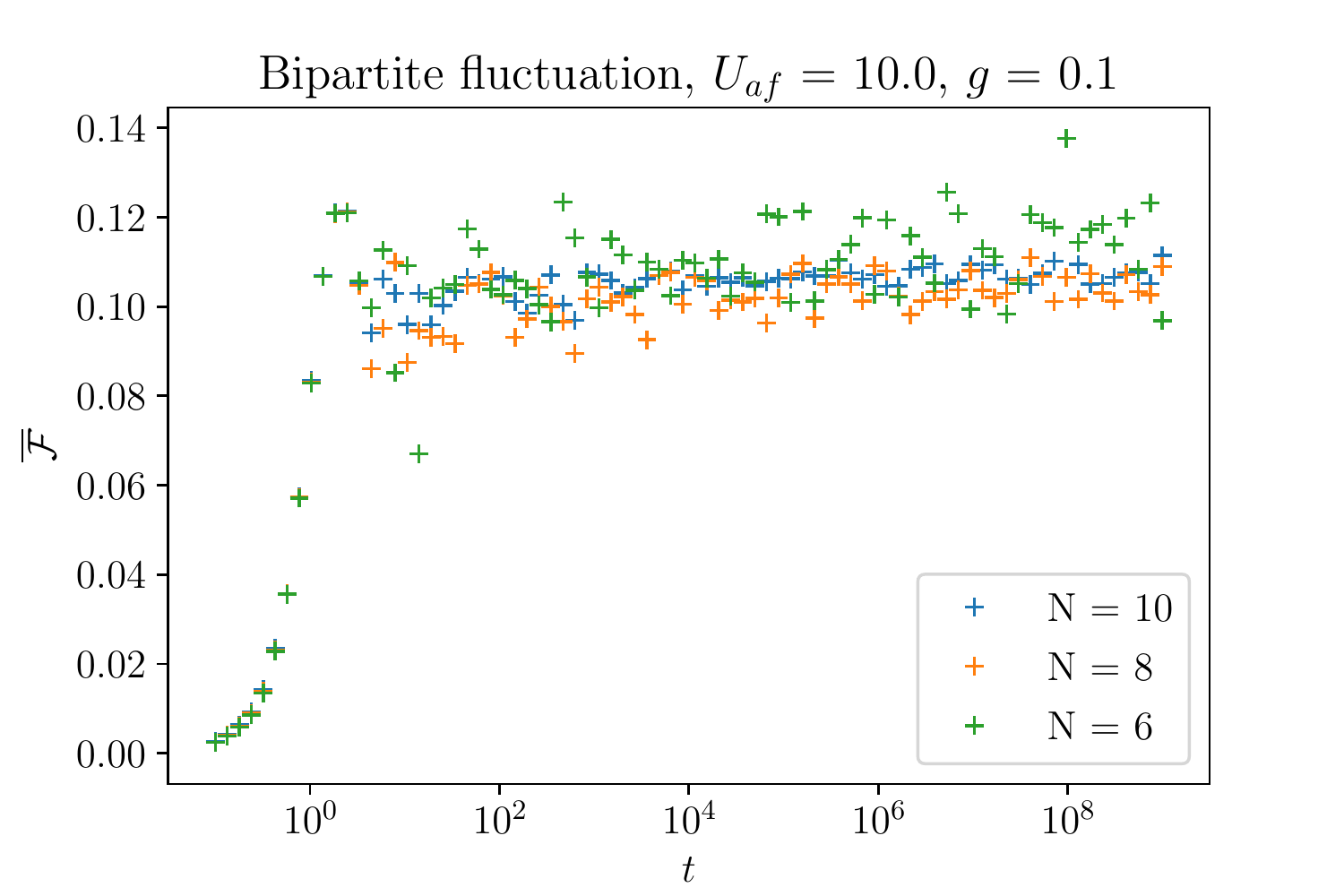}
 \caption{Entanglement entropy (left column) and bipartite fluctuation (right column) for the half chain (i.e. $l=N/2$) with $U_{af} = 1.0$ (upper line) and $U_{af} = 10.0$ (lower line). Here we show an average over all possible configurations of disorder (for number of sites $N$, there are $2^N$ possibilities) with open boundary conditions. Here we set $J_{xy}=-0.25$, $J_z = 0.25$, $g=0.1$ and $\phi=0.01$.}
\label{EE_bipartite_time}
\end{figure*}

This situation remains qualitatively similar with $g=0.1$. In Fig. \ref{EE_bipartite_time}, we show the results for the entanglement entropy (left column) and the bipartite fluctuation (right column) at $g=0.1$ with relatively weak disorder $U_{af} = 1.0$ (upper line) and with strong disorder in the localized phase with $U_{af} = 10.0$ (lower line). There we considered the time-evolution starting from the N\'eel state for all possible configurations of disorder, evaluated the bipartite fluctuation and the entanglement and consequently averaged over all possible configurations of the disorder. We consequently call the evaluated quantities
$\overline{\mathcal{F}}$ and $\overline{S}$ respectively.

\begin{figure*}
 \includegraphics[width = 0.45\linewidth]{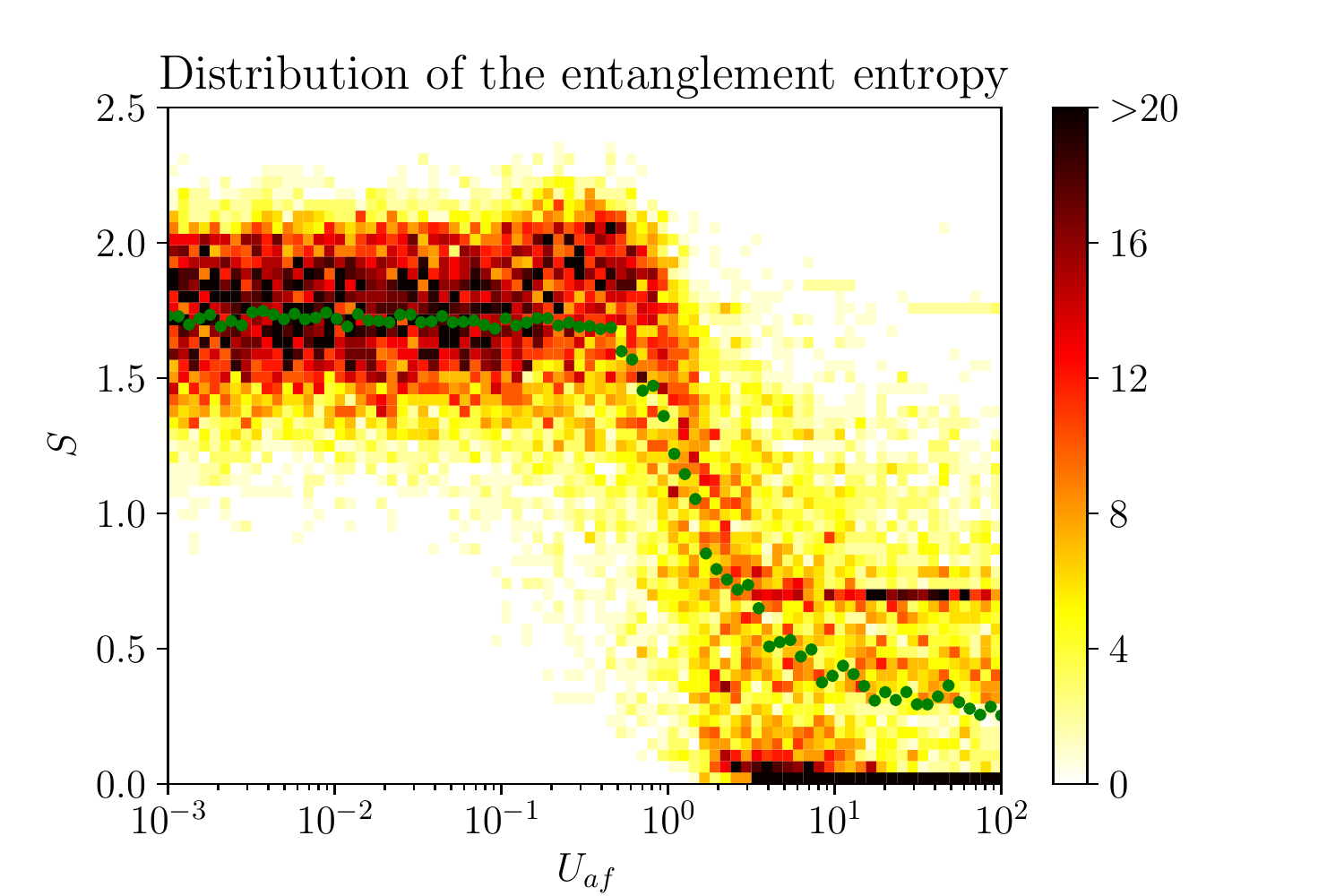}
 \includegraphics[width = 0.45\linewidth]{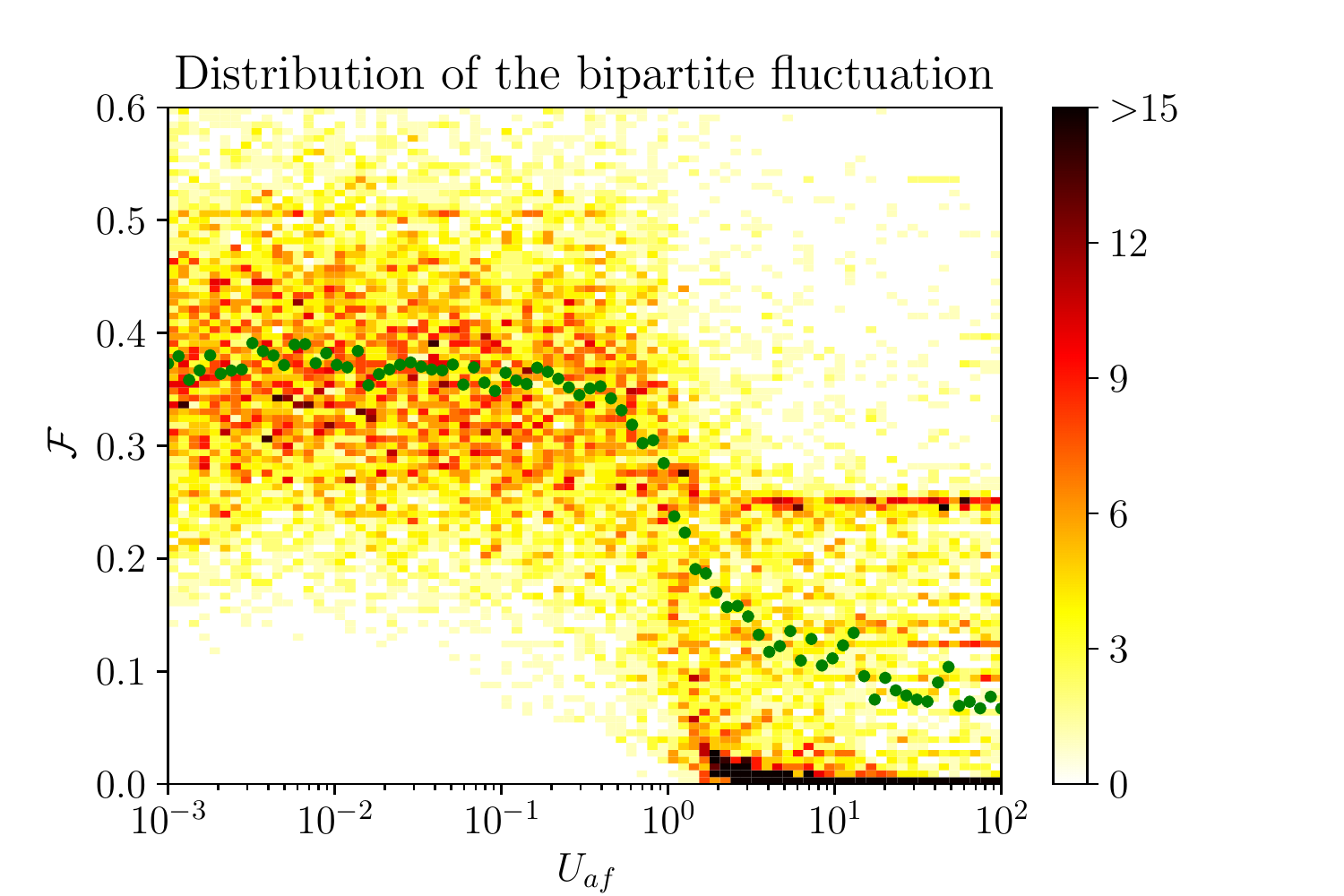}
 \caption{Distribution of the entanglement entropy (left) and the bipartite fluctuations (right) vs $U_{af}$ for $l=N/2$ and $J_{xy} = -0.25$, $J_z = 0.25$, $g=0.1$, $\phi=0.01$ and $N=8$ after $t=10^{16}$. We sampled over all $2^8 = 256$ possible configurations of the $\tau_i^z$-variables. The color scheme shows the absolute number of disorder configurations leading to a result in the respective range. The respective average $\overline{S}$ and $\overline{\mathcal{F}}$ over all disorder configurations is shown by green dots.}
 \label{EE_bipartite_long_time}
\end{figure*}

The evolution after a long time shows the localization with disorder in Fig. \ref{EE_bipartite_long_time}, which is then different from the strongly-localized regime when tracing half of the system from the ground state. In these plots we show the result for all possible configurations of disorder as a distribution in order to facilitate a comparison with the results obtained in Ref. \cite{singh2016signatures}. The color of a small square inside the plot signifies the absolute number of disorder configurations leading to such a result for the bipartite fluctuation and the entanglement entropy, respectively.
The averages over all disorder configurations $\overline{\mathcal{F}}$ and $\overline{S}$ are shown by green dots. Fig. \ref{EE_bipartite_long_time} shows the results for $g=0.1$ which are similar to the results for $g=0.0$ in a qualitative sense. We therefore conclude that the many-body localization phase is stable against a small transverse field $g$.\par

\section{Conclusion}

In this article, we have studied a bosonic ladder system populated by two different types of particles and analyzed its response to an applied magnetic field forming a $\mathbb{U}(1)$ space-dependent gauge field. We considered one particle species as impurities ({\it $f$-particles}) and analyzed the effects of their dynamics or disorder on the other boson particle species ({\it $a$-particles}). The impurities are described as two-state systems, e.g. as spinless fermions with a total density one per rung, forming a telegraph potential. In the Mott phase with one delocalized $a$-particle per rung, this model can be mapped to an effective spin model with two different types of spin-$\frac{1}{2}$ operators ($\vec{\sigma}$- and $\vec{\tau}$-spins). We assumed a density dependent coupling between $a$- and $f$-particles which gave rise to a two-spin operator $\sigma_i^z \tau_i^z$. The model and the current are invariant after a flip of all z-spin components $\sigma_i^z$ and $\tau_i^z$ (referring here to a classical Ising $\mathbb{Z}_2$ symmetry) which reflects the $1\leftrightarrow 2$ symmetry for the two legs of the ladder. The quantum impurities when moving also induce a $\mathbb{Z}_2$ gauge theory from the decoupled rungs limit and we have studied its backaction or screening effects on the $\mathbb{U}(1)$ gauge field. We have identified two distinct profiles of localization from the spin-superfluid response of the $a$-particles, i.e. a power-law form in the weakly-coupled rungs limit and a steep localization (or insulating) behaviour for strongly-coupled rungs where the current becomes strongly suppressed at a critical value of the impurities-matter coupling, and tested various forms of disorder configurations. We have formulated analytical and numerical arguments to justify these conclusions  and for the strongly-coupled rungs situation we have shown that the steep localization occurring e.g. in the antiferromagnetic case is similar to the case of fermions with a Gaussian disorder potential. For other forms of disorder configurations, the localization profile is distinct but we observe that in all studied cases the current goes to zero for the same critical value of the impurities-matter coupling which then justifies the word steep localization for all these cases assuming the strongly-coupled rungs limit. Finally, we have shown the possibility of many-body localization for the present situation of a telegraph signal formed by the impurities when deviating from a pure ground state, applying a quench and following the time evolution from a N\'eel state in the strong-interaction limit.

Our work opens further perspectives on the role of multi-particles couplings, gauge theories and many-body  localization which can be tested with current quantum technology \cite{schweizer2019floquet,googleMBL,cQEDPan}.

{\it Acknowledgements:} We acknowledge fruitful discussions with Monika Aidelsburger, Fabian Alet, Camille Aron, Valentina Ros, Thomas Baker, Fabian Grusdt, Nicolas Laflorencie, Alexandru Petrescu, Frank Pollmann, Nicolas Regnault and Stam Nicolis.
This work was funded  by the Deutsche Forschungsgemeinschaft (DFG, German Research Foundation) under Project No. 277974659 via Research Unit FOR 2414 (including the PhD thesis of EB and FY). FY recently moved to Stockholm University and acknowledges funding from the Swedish Research Council (VR) and the Knut and Alice Wallenberg Foundation. This work has also benefitted from the Hopper cluster at CPhT.

\appendix
\section{Numerical implementation}\label{numerics}

The effective spin Hamiltonians \eqref{effective_spin}, \eqref{effective_four_body} and \eqref{effective_XXZ} were implemented numerically using the QuTip package for Python \cite{johansson2012qutip}.
In the case of a static $f$-particle configuration, the Hilbert space for a chain of length $L$ is $2^L$-dimensional and the operators constituting the Hamiltonian are realized by their actual form in this Hilbert space,
\begin{equation*}
	\sigma_i^+ \sigma_{i+1}^- = \id_1 \otimes ... \otimes \id_{i-1} \otimes \sigma_i^+ \otimes \sigma_{i+1}^- \otimes \id_{i+2} \otimes ...\otimes \id_L.
\end{equation*}
In this framework, it is simple to implement the $f$-particle dynamics using quantum spins as well, in order to check Eq. \eqref{par_current_include_gf} or to explore the model \eqref{effective_four_body} featuring a four-body interaction. The Hilbert space for a chain of length $L$ is then $2^{2L}$-dimensional and the operators take the following form
\begin{widetext}
\begin{equation*}
	\sigma_i^z \tau_i^z = \id_{\sigma,1} \otimes \id_{\tau,1} \otimes ... \otimes \id_{\sigma,i-1} \otimes \id_{\tau,i-1} \otimes \sigma_i^z  \otimes \tau_i^z \otimes \id_{\sigma,i+1} \otimes \id_{\tau,i+1} \otimes ...\otimes \id_{\sigma,L} \otimes \id_{\tau,L}.
\end{equation*}
\end{widetext}
In the implemented program, once the Hamiltonian is defined, the ground state is evaluated numerically. We evaluate the expectation values of the currents by evaluating the expectation values of the spin operators and correlations in Eq. \eqref{first_currents}.\par

For certain specialized setups, we can diagonalize the Hamiltonian in momentum space after mapping to free fermions and hence evaluate the current as a sum over occupied momentum states (see Eqs. \eqref{current_sum}, \eqref{alternating_current_sum} and \eqref{current_sum_2}). In these cases, the current can be evaluated directly by evaluating occupied momenta (respecting the discretization and boundary conditions, see Sec. \ref{ferro_order_section}) and summing numerically. This procedure has then also been used to evaluate the current in the interacting case with a staggered magnetic field after bosonization. In this case, the sum was truncated to account for the gap opened by the staggered field. Note that due to the numerical simplicity of this approach, significantly larger systems can be analyzed than with the ED approach. The downside is of course that it can only be applied if a mapping to the momentum space is at all possible.\par
In Sec. \ref{RG_analysis}, we described how the sum over momentum space can be truncated in order to account for the localization with increasing disorder strength. In Eq. \eqref{fit_equation}, we introduced a fitting parameter $C$ which we evaluated numerically to make the connection between data from ED and the truncation of the sum in momentum space. The results of this procedure can be seen in Figs. \ref{localization_figure} and \ref{staggered_localization}. For the evaluation of the truncated sum in momentum space, we used a large system to approach the continuum limit, necessary for our RG analysis to hold. The ED data can be obtained only for small systems. When comparing the currents without the disorder, there is a small offset between both results which is getting larger for smaller systems in ED. In order to fit $C$ in the localization length, we neglect this offset (i.e. shift both curves onto each other).
For the fitting, we use a least square optimizer for the desired range of disorder strength to evaluate $C$. \par
In Sec. \ref{MBL}, we investigate bipartite spin fluctuations and entanglement entropy. We firstly consider both quantities in the ground state. For that, we evaluate the ground state for each possible configuration of disorder at a certain disorder strength, evaluate the bipartite fluctuation \eqref{bipartite_fluctuation} and the entanglement entropy \eqref{entanglement_entropy} by taking expectation values for each disorder configuration and finally averaged over disorder configurations. When we study the long time evolution in Sec. \ref{longTimeEvolution}, we prepare the system initially in the Néel state (for the $a$-particles) and in a certain configuration of disorder. We consider the evolved state after a time $t$ using the (time) evolution operator expressed as a matrix exponential as
\begin{equation}
  \ket{\Psi(t)} = \mathrm{e}^{-i H_\text{dis} t} \ket{\Psi(0)}.
\end{equation}
Here, we wrote $H_\text{dis}$ to underline the static disorder in the $\tau_i^z$-variables in the Hamiltonian. We then evaluate the bipartite fluctuation \eqref{bipartite_fluctuation} and the entanglement entropy \eqref{entanglement_entropy} using $\ket{\Psi(t)}$. Finally, we average over all disorder configurations.

\section{Calculations in the decoupled-rungs limit with static impurities}\label{bloch_sphere_calculation}

Here, we detail the calculations from Sec. \ref{static_impurities} which lead to Eq. \eqref{current_prediction}. To evaluate the ground state in the decoupled-rungs limit, we set $J_{xy} = J_z = 0$. Using the Bloch sphere representation \eqref{bloch_sphere}, we can write for the energy
\begin{equation}
  \label{bloch_sphere_hamiltonian}
  E = \sum_i \left( -g \cos(a'A_{\perp i} + \rho_i)\cos\Theta_i + \frac{U_{af}}{2} \tau_i^z \sin\Theta_i \right).
\end{equation}
Here, $\Theta_i$ and $\rho_i$ are defined as in Eq. \eqref{bloch_sphere} with $\Theta_i \in [-\frac{\pi}{2},\frac{\pi}{2}]$ and $\rho_i \in [0,2\pi)$. As we are considering static impurities, $\tau_i^z = \pm 1$. Minimization of the energy demands
\begin{equation*}
  \pdv{E}{\Theta_i} = \pdv{E}{\rho_i} = 0.
\end{equation*}
We obtain
\begin{align}
  \sin(a'A_{\perp i}+\rho_i) &= 0 , \label{rho_expression}\\
  \tan \Theta_i &= -\tau_i^z \frac{U_{af}}{2g}.
\end{align}
This shows why it is beneficial to choose the inclination angle as $\Theta_i \in [-\frac{\pi}{2},\frac{\pi}{2}]$ (instead of e.g. the more frequently used form with $\Theta_i \rightarrow \Theta_i+\pi/2$), since its range coincides with $\Im \arctan(x) \in (-\pi/2, \pi/2)$. We can therefore safely and without complications use the $\arctan$-operation and obtain Eq. \eqref{bloch_sphere_expressions} from
\begin{equation}
  \sin\Theta_i = \sin(\tan^{-1}\left(-\frac{U_{af}}{2g}\tau_i^z\right))
  \end{equation}
  \begin{equation}
  \cos\Theta_i = \cos(\tan^{-1}\left(-\frac{U_{af}}{2g}\tau_i^z\right)).
\end{equation}
Eqs. \eqref{bloch_sphere_expressions} designate indeed a minimum can be seen from the second derivative test with
\begin{multline}
  \pdv[2]{E}{\Theta_i} \pdv[2]{E}{\rho_i} - \left(\pdv[2]{E}{\Theta_i}{\rho_i} \right)^2 = (g \cos\Theta_i \cos(a'A_{\perp i}+\rho_i))^2 \\+ \frac{(U_{af}/2)^2}{1 + (U_{af}/(2g))^2} \cos(a'A_{\perp i}+\rho_i)
\end{multline}
From here and Eq. \eqref{rho_expression} we see that $\rho_i = -a'A_{\perp i}$ describes a minimum.\par
To calculate the current expectation value from Eq. \eqref{first_par_current}, we approximate expressions of the form $\langle \sigma_i^\alpha \sigma_{i+1}^\beta \rangle \approx \langle \sigma_i^\alpha \rangle \langle \sigma_{i+1}^\beta \rangle$ since we are considering the decoupled rung limit.
The current then reads using Eqs. \eqref{bloch_sphere}
\begin{equation*}
  \langle j_\parallel \rangle = 2{J_{xy}} \cos\Theta_i \cos\Theta_{i+1} \sin(a A_{i,i+1}^\parallel + \rho_i - \rho_{i+1}).
\end{equation*}
Plugging in the results from Eqs. \eqref{bloch_sphere_expressions}, we obtain Eq. \eqref{current_prediction}.

\section{Calculations in the decoupled-rungs limit with mobile impurities}\label{mobile_impurities_calculation}
Here we detail the calculations leading to the results presented in Sec. \ref{mobile} with mobile impurities. As described there, this corresponds to the model in Eq. \ref{effective_spin} with an additional term $-g_f \tau_i^x$ which allows the f-particles to hop between the legs of the ladder.
We assume that $g \gg {J}_{xy}, J_z$ to approximately decouple the rungs from each other. Using the notation $\ket{\up\up}_i = \ket{\up}_{\sigma_i^z}\otimes\ket{\up}_{\tau_i^z}$, in the basis $\ket{\up\up}$, $\ket{\up\down}$, $\ket{\down\up}$, $\ket{\down\down}$,
the Hamiltonian of one rung can then be written as
\begin{equation}
  H =
  \begin{pmatrix}
  U_{af}/2 & -g_f & -g \mathrm{e}^{ia'A_{\perp,i}} & 0 \\
  -g_f & -U_{af}/2 & 0 & -g\mathrm{e}^{ia'A_{\perp,i}}\\
  -g\mathrm{e}^{-ia'A_{\perp,i}} & 0 & -U_{af}/2 & -g_f\\
  0 & -g\mathrm{e}^{-ia'A_{\perp,i}} & -g_f & U_{af}/2
  \end{pmatrix}.
\end{equation}
\par If $g_f = 0$, the ground state energy $E = -\sqrt{ g^2+\left(\frac{U_{af}}{2}\right)^2}$ is twofold degenerate and is attained by the states
\begin{eqnarray}
  \ket{+} &=& N_0^+(\mathrm{e}^{ia'A_{\perp,i}}p_0^+ \ket{\up\up} + \ket{\down\up}), \nonumber \\
  \ket{-} &=& N_0^-(\mathrm{e}^{ia'A_{\perp,i}}p_0^- \ket{\up\down} + \ket{\down\down}),
\end{eqnarray}
where $p_0^{\pm} = \frac{\sqrt{4 g^2+U_{af}^2} \mp U_{af}}{2 g}$ and $N_0^\pm = 1/\sqrt{1+(p_0^\pm)^2}$. Those two states correspond to the distinction of $\tau_i^z = \pm 1$ in the case of static impurities and yield $\langle \tau_i^z \rangle = \pm 1$ respectively. The problem is then the same as in Sec. \ref{static_impurities}.
The $\sigma_i^z$-variables have eigenvalues of $\langle \sigma_i^z \rangle = -\langle \tau_i^z \rangle U_{af}/\sqrt{4g^2+U_{af}^2}$. The correlation $\langle \sigma_i^z \tau_i^z \rangle$ takes the value
\begin{equation}
  \label{correlation_expectation}
  \langle \sigma_i^z \tau_i^z \rangle = -\frac{U_{af}}{\sqrt{4g^2+U_{af}^2}}.
\end{equation}
\par
If $g_f > 0$, there is a non-degenerate ground state $N(\mathrm{e}^{ia'A_{\perp,i}}\ket{\up\up}+\ket{\down\down} + p(\mathrm{e}^{ia'A_{\perp,i}}\ket{\up\down}+\ket{\down\up}))$ with energy $E = -\frac{1}{2}\sqrt{4(g+g_f)^2+U_{af}^2}$, where
\begin{eqnarray}
  p &=& \frac{\sqrt{4 (g+g_f)^2+U_{af}^2}+U_{af}}{2 (g+g_f)}, \nonumber \\
  N &=& 1/\sqrt{2+2p^2}. \nonumber
\end{eqnarray}
In this ground state, $\sigma_i^z$ and $\tau_i^z$ now have expectation value $0$, whereas the correlation $\langle \sigma_i^z \tau_i^z \rangle$ takes the value
\begin{equation}
  \langle \sigma_i^z \tau_i^z \rangle = \frac{1-p^2}{1+p^2} = -\frac{U_{af}}{\sqrt{4(g+g_f)^2+U_{af}^2}}.
\end{equation}
This result has been confirmed by comparing to results of ED simulations which are shown in the inset of Fig. \ref{different_gf}. The expectation values of $\tau_i^z$ and $\sigma_i^z$ change to zero for $g_f > 0$, whereas they are non-zero for $g_f = 0$.
We also identify $\langle \tau_i^x\rangle = \langle {\cal I}\otimes \tau_i^x\rangle = \frac{2p}{1+p^2}$ and ${\cal I}$ is the identity operator or $2\times 2$ identity matrix acting on the Hilbert space of the $\vec{\sigma}$-spin.
\par
In a similar way, the expectation values of $\sigma_i^x$ and $\sigma_i^y$ can be calculated. Invoking a mean field approach and plugging the results into Eq. \eqref{first_currents}, the result in Eq. \eqref{par_current_include_gf} can be found. Eq. \eqref{par_current_include_gf} goes to eq. \eqref{current_prediction} for $g_f \rightarrow 0$, so it is continuous in $g_f$.

\section{Derivation of Four-Body Hamiltonian}\label{fourbody_derivation}

As described in Sec. \ref{large_impurity_interaction}, if we include the hopping of the impurities in all directions, we have to enhance Hamiltonian \eqref{petrescu_hamilton} by these new processes and potentials constraining this mobility.
We therefore consider the following Hamiltonian:
\begin{widetext}
\begin{multline}
\label{enhanced_petrescu_hamilton}
  H = -t^a_x \sum_{\alpha,i} \mathrm{e}^{iaA_{i,i+1}^\alpha} a^{\dag}_{\alpha i} a_{\alpha,i+1 } +\text{h.c.}
  - t^a_y \sum_{i} \mathrm{e}^{-ia'A_{\perp i}} a^{\dag}_{2 i} a_{1 i}  - t_x^f \sum_{\alpha,i} f^\dag_{\alpha i} f_{\alpha,i+1 }- t^f_y \sum_{i} f^{\dag}_{2 i} f_{1 i} + \text{h.c.} \\
   + \frac{U_{aa}}{2}\sum_{\alpha,i} n^a_{\alpha i} (n^a_{\alpha i}-1) + \frac{U_{ff}}{2}\sum_{\alpha,i} n^f_{\alpha i} (n^f_{\alpha i}-1) + V_{\perp}\sum_{ i} (n^a_{1 i} n^a_{2 i} + n^f_{1 i} n^f_{2 i})- \mu \sum_{\alpha, i} n^a_{\alpha i} + U_{af} \sum_{\alpha,i} n_{\alpha,i}^a n_{\alpha,i}^f.
\end{multline}
\end{widetext}
With respect to the setup \eqref{petrescu_hamilton} where the impurities were considered as static, we added the hopping of the $f$-particles in x- and y-direction with the amplitudes $t^f_x$ and $t^f_y$ respectively. Furthermore, we added potentials penalising two $f$-particles sitting on the same site ($U_{ff}$) and on the same rung ($V_\perp$) in completely analogous to the $a$-particle dynamics in Sec. \ref{TheModel}. While the on-site repulsion ($U_{ff}$) can be chosen freely, we constrain $V_\perp$ for simplicity to be the same for both $a$- and $f$-particles.
In the following, we will consider only the limit where formally $U_{ff} \rightarrow \infty$ is the biggest parameter and therefore two impurities cannot occupy the same site. They thus behave like spinless fermions.
As in Sec. \ref{TheModel}, we aim to derive an effective spin model at half-filling of both particle species while now considering $U_{af}$ of the same order as $U_{aa}$ and $V_\perp$ and much larger than the hopping amplitudes.
The ground state without hopping is now that with an $a$- and an $f$-particle on opposite legs on each rung, there is a $\mathbb{Z}_2$ gauge freedom on each rung. Restoring the hopping along the legs perturbatively, it gives rise to second order processes for both $a$-and $f$-particles in an analogous way as in Sec. \ref{TheModel} which can be written as
\begin{equation}
  J_z^a \sigma_i^z\sigma_{i+1}^z + J_z^f \tau_i^z\tau_{i+1}^z
\end{equation}
with
\begin{eqnarray}
  J_z^a &=& (t_x^a)^2 \left( \frac{2}{U_{aa}}-\frac{1}{V_\perp+U_{af}} \right), \nonumber \\
  J_z^f &=& -\frac{(t_x^f)^2}{V_\perp+U_{af}}.
\end{eqnarray}
Due to the mobility of the impurities and the large inter-species interaction $U_{af}$, the hopping along the rungs also enters through a second order process which accounts for exchange of the two species along a rung which reads
\begin{equation}\label{perturbation_Josephson_term}
  -g^{af}\mathrm{e}^{ia'A_{\perp i}} \sigma_i^+ \tau_i^- +\text{h.c}.
\end{equation}
with
\begin{equation}
  g^{af} = 2\frac{t_y^a t_y^f}{U_{af}}.
\end{equation}
To evaluate the current along the legs, processes need to be considered which interchange the $a$-particle states between two neighboring rungs. Due to the large interspecies potential $U_{af}$, these interchange also the states of the $f$-particles so that the relevant processes interchange completely the state between two rungs. Such processes arise to fourth order in perturbation theory for neighbouring rungs with initially different configuration. Therefore, all of the resulting terms contain either the four-body operator $\sigma_i^-\tau_i^+\sigma_{i+1}^+\tau_{i+1}^-$ or its Hermitian conjugate.
We can distinguish such processes where the $a$-particles hop only along the legs (one of them is shown exemplarily in Fig. \ref{fourth_order_exchange}) which bear a phase factor $\mathrm{e}^{ia(A^1_{i,i+1}-A^2_{i,i+1})}$ ot its Hermitian conjugate, and such processes where the $a$-particles hop only along the rungs, therefore having a phase factor $\mathrm{e}^{ia'(A_{\perp i}-A_{\perp i+1})}$ or its Hermitian conjugate.
Note that we consider a setup where the $f$-particles are not affected by the magnetic field and therefore do not acquire a phase upon hopping. Altogether and with our assumptions we can identify 88 fourth order processes leading to an exchange of two initially different configurations between neighbouring rungs on a plaquette.
We can write the arising term as
\begin{widetext}
\begin{equation}\label{full_fourth_order_term}
  -(J_{xy}^\parallel \mathrm{e}^{-ia(A^1_{i,i+1}-A^2_{i,i+1})} + J_{xy}^\perp \mathrm{e}^{-ia'(A_{\perp i}-A_{\perp i+1})})\sigma_i^-\tau_i^+\sigma_{i+1}^+\tau_{i+1}^-+ \text{h.c.} ,
\end{equation}
with
\begin{eqnarray}\label{Jxy_parallel}
  J_{xy}^\parallel &=&  \frac{8(t_x^a)^2 (t_y^f)^2}{U_{af}+V_\perp}\left(\frac{1}{U_{af}^2} + \frac{1}{U_{af}(U_{af}+V_\perp)} + \frac{1}{2(U_{af}+V_\perp)^2} \right)  + \frac{8(t_x^a)^2 (t_x^f)^2 }{(U_{af}+V_\perp)^2} \left( \frac{1}{2 U_{af}} + \frac{1}{2 V_\perp} + \frac{1}{2 U_{af}+2 V_\perp}\right),
  \nonumber \\
  J_{xy}^\perp &=& 8 \frac{(t_y^a)^2 (t_y^f)^2}{U_{af}^3}   + 8 \frac{(t_y^a)^2 (t_x^f)^2 }{U_{af}+V_\perp}\left(\frac{1}{U_{af}^2} + \frac{1}{U_{af}(U_{af}+V_\perp)} + \frac{1}{2(U_{af}+V_\perp)^2} \right).
\end{eqnarray}
\end{widetext}
As we are mainly interested in the parallel current and we assume that the term \eqref{perturbation_Josephson_term} is dominant, the terms proportional to $ J_{xy}^\perp$ in \eqref{full_fourth_order_term} can be neglected as in that case as the total phase vanishes there. We therefore consider only the term
\begin{equation}\label{simplified_fourth_order_term}
  -J_{xy}^\parallel \mathrm{e}^{-ia(A^1_{i,i+1}-A^2_{i,i+1})}\sigma_i^-\tau_i^+\sigma_{i+1}^+\tau_{i+1}^-+ \text{h.c}.
\end{equation}

\bibliography{stage_biblio}

\end{document}